\documentclass[%
 aip,
 amsmath,amssymb,
 reprint,%
]{revtex4-2}

\usepackage{graphicx}% Include figure files
\usepackage{dcolumn}% Align table columns on decimal point
\usepackage{bm}% bold math
\usepackage[utf8]{inputenc}
\usepackage[T1]{fontenc}
\usepackage{mathptmx}
\usepackage{etoolbox}
\usepackage{scalefnt}

\makeatletter
\def\@email#1#2{%
 \endgroup
 \patchcmd{\titleblock@produce}
  {\frontmatter@RRAPformat}
  {\frontmatter@RRAPformat{\produce@RRAP{*#1\href{mailto:#2}{#2}}}\frontmatter@RRAPformat}
  {}{}
}%
\makeatother
\begin{document}

\preprint{AIP/123-QED}

\title[Collective transport of droplets through porous media]{Collective transport of droplets through porous media}
% Force line breaks with \\

    \author{Rodrigo C. V. Coelho}
    \email{rcvcoelho@fc.ul.pt}
	\affiliation{Centro de F\'isica Te\'orica e Computacional, Faculdade de Ciências, Universidade de Lisboa,
		P-1749-016 Lisboa, Portugal}
    \affiliation{Departamento de F\'{\i}sica, Faculdade
		de Ci\^{e}ncias, Universidade de Lisboa, P-1749-016 Lisboa, Portugal}
	
	\author{Danilo P. F. Silva}
	\affiliation{Centro de F\'isica Te\'orica e Computacional, Faculdade de Ciências, Universidade de Lisboa,
		P-1749-016 Lisboa, Portugal}
    \affiliation{Departamento de F\'{\i}sica, Faculdade
		de Ci\^{e}ncias, Universidade de Lisboa, P-1749-016 Lisboa, Portugal}
	
	\author{António M. R. Maschio}
	\affiliation{Centro de F\'isica Te\'orica e Computacional, Faculdade de Ciências, Universidade de Lisboa,
		P-1749-016 Lisboa, Portugal}
    \affiliation{Departamento de F\'{\i}sica, Faculdade
		de Ci\^{e}ncias, Universidade de Lisboa, P-1749-016 Lisboa, Portugal}
	
	\author{Margarida M. Telo da Gama}
	\affiliation{Centro de F\'isica Te\'orica e Computacional, Faculdade de Ciências, Universidade de Lisboa,
		P-1749-016 Lisboa, Portugal}
    \affiliation{Departamento de F\'{\i}sica, Faculdade
		de Ci\^{e}ncias, Universidade de Lisboa, P-1749-016 Lisboa, Portugal}
	
	\author{Nuno A. M. Ara\'ujo}
	\affiliation{Centro de F\'isica Te\'orica e Computacional, Faculdade de Ciências, Universidade de Lisboa,
		P-1749-016 Lisboa, Portugal}
    \affiliation{Departamento de F\'{\i}sica, Faculdade
		de Ci\^{e}ncias, Universidade de Lisboa, P-1749-016 Lisboa, Portugal}

\date{\today}% It is always \today, today,

\begin{abstract}
The flow of deformable particles, such as droplets, dragged by a fluid, through a network of narrow pores inside rocks or other porous media is key in a range of applications, from enhanced oil recovery and water filtration to lab on a chip sorting of cells. The collective dynamics and its impact on the flow are poorly understood. Here, using droplets as a prototype, we show that collective transport can occur for conditions under which a single particle would get trapped at a pore channel. When a series of droplets gets trapped, the fluids flow is affected significantly, leading to an increase of the pressure difference across the pore channels, which in turn squeezes the particles through the channels. We analyze the conditions for a single droplet to flow through one pore and derive the corresponding Bond number. We also obtain a rule for the collective flow of droplets in porous media.
\end{abstract}

\maketitle

\section{Introduction}

The transport of deformable particles and droplets through a porous material underlies a wealth of applications and natural processes. 
A prominent example is enhanced oil extraction. The standard technique for extraction is through water flooding. Because of the reservoir heterogeneity and the high viscosity of the oil, it is estimated that this technique can extract only up to 30\% of the oil contained in the reservoir~\cite{Kamal2015}. Most of the oil remains in trapped fluid volumes and as droplets blocking the narrower pores. Many techniques have been proposed to overcome this challenge and improve the  oil recovery efficiency. One of these consists in injecting water soluble polymers~\cite{Xia2008, WEVER20111558, Han1999}. The interplay between elongation and relaxation of the polymers as they are advected through the pores leads to unusual flow behaviors that promote the extraction of trapped volumes of oil in the reservoir~\cite{browne_shih_datta_2020, https://doi.org/10.1002/smll.201903944, D0SM00390E}. A second promising approach relies on the injection of deformable micro-gel particles into the porous rock. The particles squeeze through the pores and eventually clog regions of high permeability forcing the fluid to flow through the low permeability regions~\cite{https://doi.org/10.1002/ese3.563, Bai2007, Bai2007-2}. Driven by applications such as these, a large number of studies have focused on the deformation of particles in pores~\cite{doi:10.1063/1.5139887, Bai2007}. However, the focus has been mainly on single particles and thus, questions such as how the collective transport depends on the density of flexible particles are still open.

When flexible particles are densely packed they behave collectively and the flow may be described as a complex fluid with effective properties, distinct from those of the single phase fluid~\cite{PERAZZO2018305}. For droplets, it was shown numerically that this liquid-liquid emulsion exhibits non-newtonian behaviour, namely shear thinning, when confined in a channel~\cite{arxiv.2101.06981,  C9SM02331C, PhysRevLett.119.208002}. These emulsions may self-organize in divergent microfluidic channels~\cite{PhysRevFluids.6.023606} or break-up due to the interactions between the droplets~\cite{doi:10.1063/5.0057501}. In this case, the collective effects are more intuitive as they result from the contact between the particles. At lower densities, it is not obvious how they interact at a distance through hydrodynamics and behave collectively in a complex geometry. 

Droplets in fluid flow may exhibit collective dynamics driven by the hydrodynamic interactions. For example, in a microfluidics device they exhibit vibrational modes analogous to acoustic phonons~\cite{Beatus2006}. Also flexible particles in general may show unexpected behavior. A recent microfluidics experiment suggests that deformable particles may cooperate through hydrodynamics to flow through constrictions, where, under the same conditions, one single particle is trapped~\cite{C9SM00300B}. This stems from pairwise hydrodynamic interactions: when the first particle is trapped at the constriction, the fluid velocity around it increases enabling the second particle to squeeze through the same constriction if the dimensions are appropriate. This mechanism may be used to sort particles by flexibility and size as the larger particles or less flexible ones are slowed down. Although this study addresses the hydrodynamic interactions of particles in a single pore, their interaction in porous media (many pores) remains to be investigated.

In this paper, we address the hydrodynamic interactions of many droplets in porous media. We considered two fluids (droplets and surrounding fluid) with the same viscosity. The droplets pass though a porous medium composed of a regular lattice of circular hydrophobic obstacles. We use a numerical model based on the lattice Boltzamnn method (LBM) of droplets with frustrated coalescence~\cite{arxiv.2101.06981}. We investigate the conditions for which a single droplet flows through a constriction and generalize them to porous media.
 We show that the droplets exhibit collective dynamics that allow them to flow through obstacles under conditions for which a single droplet is trapped. 
 
This paper is organized as follows. In Sec.~\ref{method-sec}, we describe the numerical method used to simulate the droplets. In Sec.~\ref{lattice-sec}, we analyse the collective dynamics of flowing droplets in porous media. We also investigate the mechanism through which the droplets cooperate and the conditions for which a single droplet flows through a constriction. Finally, in Sec.~\ref{conclusion.sec}, we summarize the main findings and draw some conclusions.

\section{Method}
\label{method-sec}

We implemented a multicomponent pseudopotential LBM~\cite{arxiv.2101.06981} to simulate the flow of immiscible droplets in a porous medium. The motion of the two fluid components (droplets and surrounding fluid) is described by a set of distribution functions $f_{k, \alpha}(\boldsymbol{x}, t)$ at position $\boldsymbol{x}$ and time $t$ where the subscripts $k$ and $\alpha$ denote the fluid component and discrete velocity, respectively. The lattice Boltzmann equation with a force term $\mathcal{F}_{k, \alpha}$ is given by
\begin{equation}
 \begin{aligned}
 &f_{k, \alpha}\left(\boldsymbol{x}+\boldsymbol{\xi}_{\alpha} \delta t, t+\delta t\right)-f_{k, \alpha}\left(\boldsymbol{x}, t\right)=\\
 &  -\frac{\delta t}{\tau_k}\left[f_{k, \alpha}\left(\boldsymbol{x}, t\right)-f_{k, \alpha}^{eq}\left(\boldsymbol{x}, t\right)\right] + \mathcal{F}_{k, \alpha},
 \end{aligned}
 \label{eq:LBE}
 \end{equation}
 where $\boldsymbol{\xi}_{\alpha}$ is the velocity vector, $\tau_k$ is the relaxation time and $f_{k, \alpha}^{eq}\left(\boldsymbol{x}, t\right)$ is the distribution at equilibrium. $\delta x$ and $\delta t$ represent the physical distance between two adjacent lattice nodes and the time step.  We express the results in lattice units (l.u.) i.e. $\delta x=1$, $\delta t=1$. The fluid viscosity $\nu_k$ is related to $\tau_k$ as:
\begin{equation}
\nu_k=c_{s}^{2}\left(\tau_k-\frac{1}{2}\right).
\label{eqn:viscosity_ls}
\end{equation}
where $c_{s}$ is the speed of sound. The equilibrium distribution $f_{k, \alpha}^{eq}$ depends on the macroscopic fluid velocity and density:
\begin{equation}
f_{k, \alpha}^{eq}\left(\boldsymbol{x}, t\right)=w_{\alpha} \rho_k \left[1+\frac{\boldsymbol{\xi}_{\alpha} \cdot \boldsymbol{u}^{eq}}{c_{s}^{2}}+\frac{\left(\boldsymbol{\xi}_{\alpha} \cdot \boldsymbol{u}^{eq}\right)^{2}}{2 c_{s}^{4}}-\frac{(\boldsymbol{u}^{eq})^{2}}{2 c_{s}^{2}}\right],
\end{equation}
where $\boldsymbol{u}^{eq}$ is an effective velocity and $w_{\alpha}$ are the lattice weights. For the lattice, we considered the D3Q41 in the streaming step (see the Appendix). 
The effective velocity arises from assuming that, in the absence of interparticle interactions, the equilibrium velocities for each of the $k$th fluid components are equal to a common effective velocity $\boldsymbol{u}^{eq}$, which is given by
\begin{equation}
\boldsymbol{u}^{e q}=\sum_{k} \frac{\rho_{k} \boldsymbol{u}_{k}}{\tau_{k}} / \sum_{k} \frac{\rho_{k}}{\tau_{k}},
\end{equation}
where $\rho_{k}$ and $\boldsymbol{u}_{k}$ are the density and velocity of the $k$th fluid component, respectively, which are obtained through the following expressions
\begin{equation}
\begin{aligned}
\rho_k &=\sum_{\alpha} f_{k, \alpha},\\
\rho_k \boldsymbol{u}_k &=\sum_{\alpha} \boldsymbol{\xi}_{\alpha} f_{k, \alpha} + \frac{\boldsymbol{F}_k}{2}.
\end{aligned}
\end{equation}
The barycentric velocity and total density of the fluid mixture are given by
\begin{equation}
\boldsymbol{u}=\frac{\sum_{k} \rho_{k} \boldsymbol{u}_{k}}{\rho}, \quad \rho=\sum_{k} \rho_{k}.
\end{equation}
The force term $\mathcal{F}_{k, \alpha}$ is introduced to model external force fields, such as gravity or forces between the two different components \cite{Zhao_Li_2002},
\begin{equation}
    \mathcal{F}_{k, \alpha}=\left(1-\frac{1}{2\tau_k}\right) w_{\alpha}\left(\frac{\boldsymbol{\xi}_{\alpha}-\boldsymbol{u}^{eq}}{c_{s}^{2}}+\frac{\boldsymbol{\xi}_{\alpha} \cdot \boldsymbol{u}^{eq}}{c_{s}^{4}} \boldsymbol{\xi}_{\alpha}\right) \cdot \boldsymbol{F}_k.
\end{equation}

The total force $\boldsymbol{F}_k$ acting on fluid component $k$ is a combination of three forces: a repulsive force $\boldsymbol{F}_{k}$ between the fluid components to mimic fluid immiscibility, an attractive force (1st and 3rd belts), and a repulsive force (first three belts) $\boldsymbol{F}_{k}^{c}$. The latter two competing forces are required to prevent coalescence of the droplets. Then 
\begin{equation}
\boldsymbol{F}_{k}=\boldsymbol{F}_{k}^{r}+\boldsymbol{F}_{k}^{c},
\end{equation}
where  $\boldsymbol{F}^r_{k}$ is determined through~\cite{kruger2016} 
\begin{equation}
\boldsymbol{F}_{k}^{r}=-\rho_{k}(\boldsymbol{x}) \sum_{\bar{k}} G_{\bar{k} k} \sum_{\alpha=0}^{40} w_{\alpha} \rho_{\bar k}\left(\boldsymbol{x}+\boldsymbol{\xi}_{\alpha}\right)\boldsymbol{\xi}_{\alpha}.
\end{equation}
$G_{\bar{k} k}$ is the parameter that sets the strength of the interaction between the fluids. The pseudopotential is  $\psi_{k}=\rho_{\bar{k}}$ where the bar over $k$ indicates the other fluid component. For simplicity, we set $G_{\bar{k} k}= G_{k \bar{k}}$. To achieve fluid-fluid separation we set $G_{k, \bar{k}}>0$. $\boldsymbol{F}_{k}^{c}$ is determined through
\begin{equation}
\begin{split}
    \boldsymbol{ F } _ { k } ^ { c } = - G _ { k , 1 } \psi _ { k } ( \boldsymbol{ x } ) \sum _ { \alpha = 0 } ^ { 40 } w _ { \alpha } \psi _ { k } \left( \boldsymbol { x } + \boldsymbol {\xi}_{\alpha} \right) \boldsymbol{\xi}_{\alpha}\\ - G _ { k , 2 } \psi _ { k } ( \boldsymbol { x } ) \sum _ { \beta = 0 } ^ { 38 } w _ { \beta } \psi _ { k } \left( \boldsymbol { x } + \boldsymbol {\xi}_{\beta} \right) \boldsymbol {\xi}_{\beta},
    \end{split}
    \label{eq:competing_force}
\end{equation}
where the first term corresponds to the attractive and the second to the repulsive  forces. In the implementation of the second term, we considered a different lattice velocity, the D3Q39, and thus the subscript $\beta$ refers to the D3Q39 and $\alpha$ to D3Q41 lattices (see the Appendix for the details of these lattices). The pseudopotential is $\psi _ { k } = \rho _ { 0 } \left( 1 - e ^ { - \rho _ { k } / \rho _ { 0 } } \right)$ with a uniform reference density $\rho _ { 0 }$ and  $G _ { k , 1 }$ and $ G _ { k , 2 }$ are the self-interaction strength coefficients. Again, for simplicity we choose $G _ { k , 1 } = G _ { \bar{k} , 1 }$ and $G _ { k , 2 } = G _ { \bar{k} , 2 }$. To prevent coalesce, we set $G_{k, 1}<0$ and $G_{k, 2}>0$ with $\left|G_{k, 1}\right|>\left|G_{k, 2}\right|$. The values used in the simulations of the droplets are $G_{\bar{k} k}=3$, $G_{k, 1}=-7.9$ and $G_{k, 2}=4.9$.
The equation of state (EOS) is given by
\begin{align}
& P= \\
&\sum_{k=k, \bar{k}}\left[\rho_{k} c_{s}^{2}+\frac{1}{2} G_{k, 1} c_{s}^{2} \psi_{k}^{2}+\frac{1}{2} G_{k, 2} c_{s}^{2} \psi_{k}^{2}+\frac{1}{2} G_{k \bar{k}} \psi_{k} \psi_{\bar{k}} c_{s}^{2}\right] .
\label{pressure-eq}
\end{align}

\begin{figure}
	\centering
	\includegraphics[width=\columnwidth]{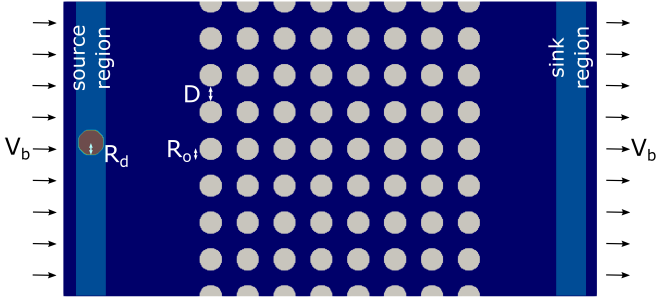}
	\caption{Scheme of the simulation setup with the relevant dimensions: $R_d$ is the droplet radius, $R_o$ is the radius of the circular obstacles and $D$ is the pore aperture.}\label{fig-scheme}
\end{figure}

We impose inflow-outflow boundary conditions with fixed velocity $v_b$ at the inlet and fixed density $\rho_{out}$ at the outlet through a nonequilibrium extrapolation approach which is second order accurate \cite{Zhao_Li_2002}. The boundary conditions are implemented as post-collision distribution functions $\tilde{f}_{k, \alpha}(\boldsymbol{x_b}, t)$ at the boundary nodes
\begin{equation}
\tilde{f}_{k, \alpha}(\boldsymbol{x_b})=f_{k, \alpha}^{e q}(\boldsymbol{x_b})+(1-\omega_k) f_{k, \alpha}^{n e q}(\boldsymbol{x_b}),
\end{equation}
where $\boldsymbol{x_b}$ stands for a boundary node and $\omega_k= 1/\tau_k$. We take for the nonequilibrium distribution $f_{k, \alpha}^{n e q}(\boldsymbol{x_b})$ a first order approximation from the neighbouring fluid as 
\begin{equation}
f_{k, \alpha}^{n e q}(\boldsymbol{x_b})=f_{k, \alpha}(\boldsymbol{x_f}) - f_{k, \alpha}^{e q}(\boldsymbol{x_f}) ,
\end{equation}
where $\boldsymbol{x_f}$ is the neighbouring fluid node. For the velocity boundary condition, since the velocity is known but the density is not, to obtain $f_{k, \alpha}^{e q}(\boldsymbol{x_b})$  we can extrapolate the density from the neighbouring fluid node as
\begin{equation}
\begin{aligned}
&f_{k, \alpha}^{eq}\left(\boldsymbol{x}_{b}\right)= w_{\alpha} \rho_k \left(\boldsymbol{x}_{f}\right) \\ & \left[1+\frac{\boldsymbol{\xi}_{\alpha} \cdot \boldsymbol{u}^{eq}\left(\boldsymbol{x}_{b}\right)}{c_{s}^{2}}+\frac{\left(\boldsymbol{\xi}_{\alpha} \cdot \boldsymbol{u}^{eq}\left(\boldsymbol{x}_{b}\right)\right)^{2}}{2 c_{s}^{4}}-\frac{(\boldsymbol{u}^{eq}\left(\boldsymbol{x}_{b}\right))^{2}}{2 c_{s}^{2}}\right].
\end{aligned}
\end{equation}
For the pressure boundary condition, the pressure (density) is known but the velocity is not. We extrapolate the velocity from the neighbouring fluid node as
\begin{equation}
\begin{aligned}
&f_{k, \alpha}^{eq}\left(\boldsymbol{x}_{b}\right)= w_{\alpha} \rho_k\left(\boldsymbol{x}_{b}\right) \\ & \left[1+\frac{\boldsymbol{\xi}_{\alpha} \cdot \boldsymbol{u}^{eq}\left(\boldsymbol{x}_{f}\right)}{c_{s}^{2}}+\frac{\left(\boldsymbol{\xi}_{\alpha} \cdot \boldsymbol{u}^{eq}\left(\boldsymbol{x}_{f}\right)\right)^{2}}{2 c_{s}^{4}}-\frac{(\boldsymbol{u}^{eq}\left(\boldsymbol{x}_{f}\right))^{2}}{2 c_{s}^{2}}\right].
\end{aligned}
\end{equation}
After obtaining the equilibrium and the non-equilibrium distributions, the post-collision distributions may be computed.

We impose no-slip boundaries at the surface of the obstacles. We implement these using the half-way bounce back scheme. This consists in reflecting the distribution functions as $f_{k, \bar{\alpha}}\left(\boldsymbol{x}_{b}, t+\delta t\right) = \tilde{f}_{k, \alpha}\left(\boldsymbol{x}_{b}, t\right)$. Here $\bar{\alpha}$ represents the opposite direction to $\alpha$. At the top and bottom boundaries, we impose periodic boundaries. Finally, we impose non-wetting boundary conditions between the droplets and the solid nodes by setting their virtual solid density as being that of the surrounding fluid~\cite{https://doi.org/10.1002/fld.4988}.

\begin{figure}
	\centering
	\includegraphics[width=\columnwidth]{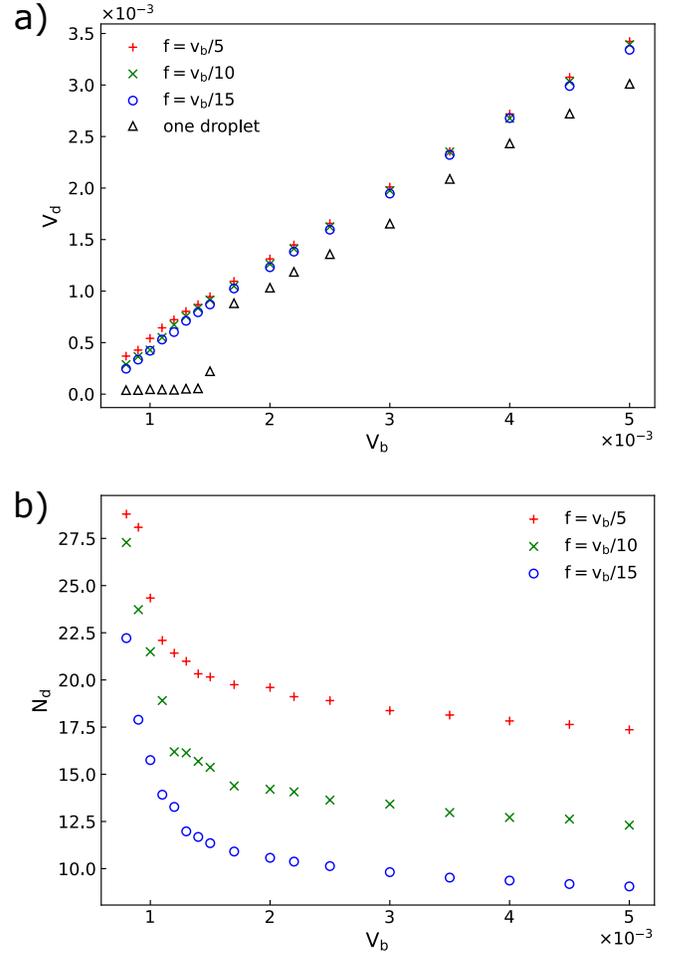}
	\caption{a) Average droplet velocity and b) average number of droplets in the medium as a function of the boundary velocity for three frequencies of droplet addition. The averages are in time and space.}\label{fig-vel-num}
\end{figure}
\begin{figure}
	\centering
	\includegraphics[width=\columnwidth]{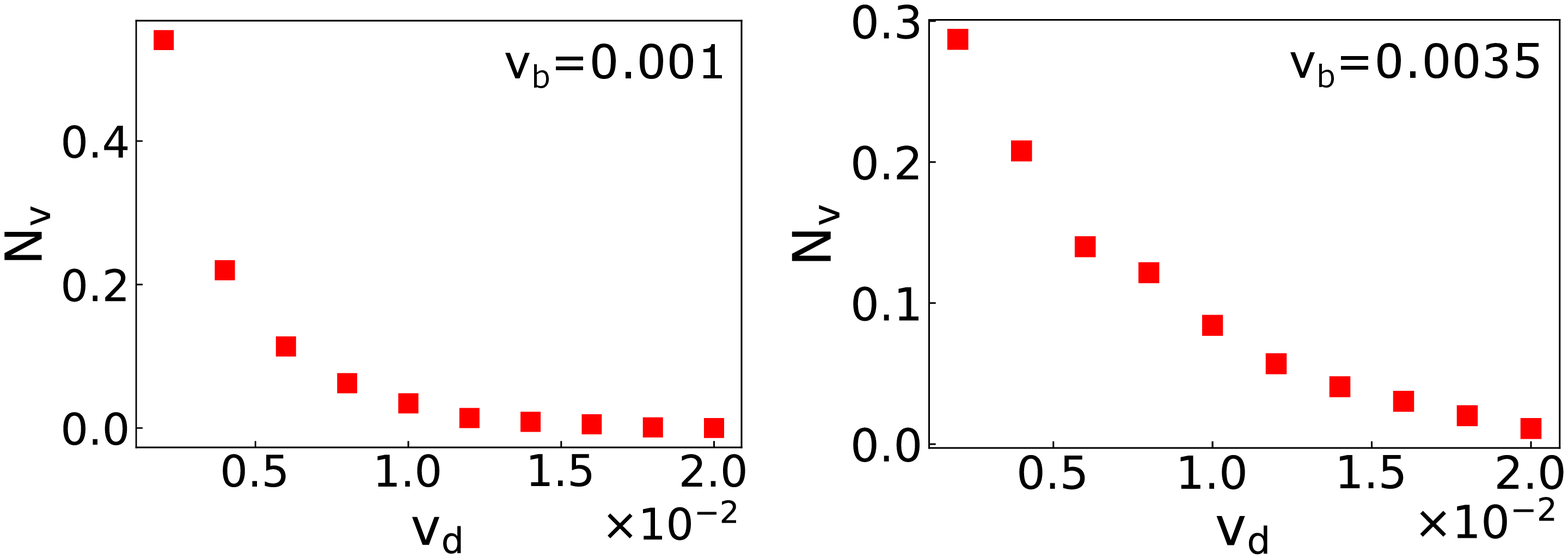}
	\caption{Histograms of the velocity of the droplets for two boundary velocities. $N_v$ is the number of counts of the droplet velocity $V_d$ divided by the total number of counts.}\label{fig-hist}
\end{figure}

\section{Results}
\label{lattice-sec}

We consider the flow through a domain with circular obstacles of radius $R_o=13$, distributed in a $8\times 8$ square lattice arrangement with a minimum distance between the surface of the obstacles $D=14$ in both directions as show in Fig.~\ref{fig-scheme}. The dimensions of the simulation domain are $L_X \times L_Y=580 \times 321$, $\rho_0=1$, $\tau_A=\tau_B=1$ and the initial densities of the two components are $\rho_A=1.22$ and $\rho_B=0.035$ inside the droplets and $\rho_A=0.035$ and $\rho_B=1.22$ outside. Periodic conditions apply in the y-direction (perpendicular to the flow) while a constant velocity $v_b$ is imposed on the left and the boundary is open on the right, i.e, zero gradient in the velocity field. Since the simulations are effectively in two dimensions, we set $L_z=1$ and apply periodic conditions in this direction. There is a source region on the left where droplets of radius $R_d=14$ are added at a certain rate and at random heights, avoiding overlaps between the created and the existing droplets. There is also a sink region on the right where droplets are removed: we reset the densities in this region at every $500$ iterations. In the source and sink regions, the density is imposed while the velocity is unchanged, which is achieved by setting the distribution function as the equilibrium one for the given fields. Note that the treatment in these regions is different from that of the inlet and outlet boundary conditions. 

Let us start with the simpler case of only one droplet in the medium. From Fig.~\ref{fig-vel-num}a, one can observe that the droplet are trapped at low velocities. This happens because the surrounding fluid can still flow through the remaining pores that are not blocked by the droplet. As $v_b$ increases beyond a threshold value  $v_b\approx 0.0017$, the droplet flow through. At higher velocities, the droplet velocity increases linearly with the boundary velocity.

\begin{figure}
	\centering
	\includegraphics[width=\columnwidth]{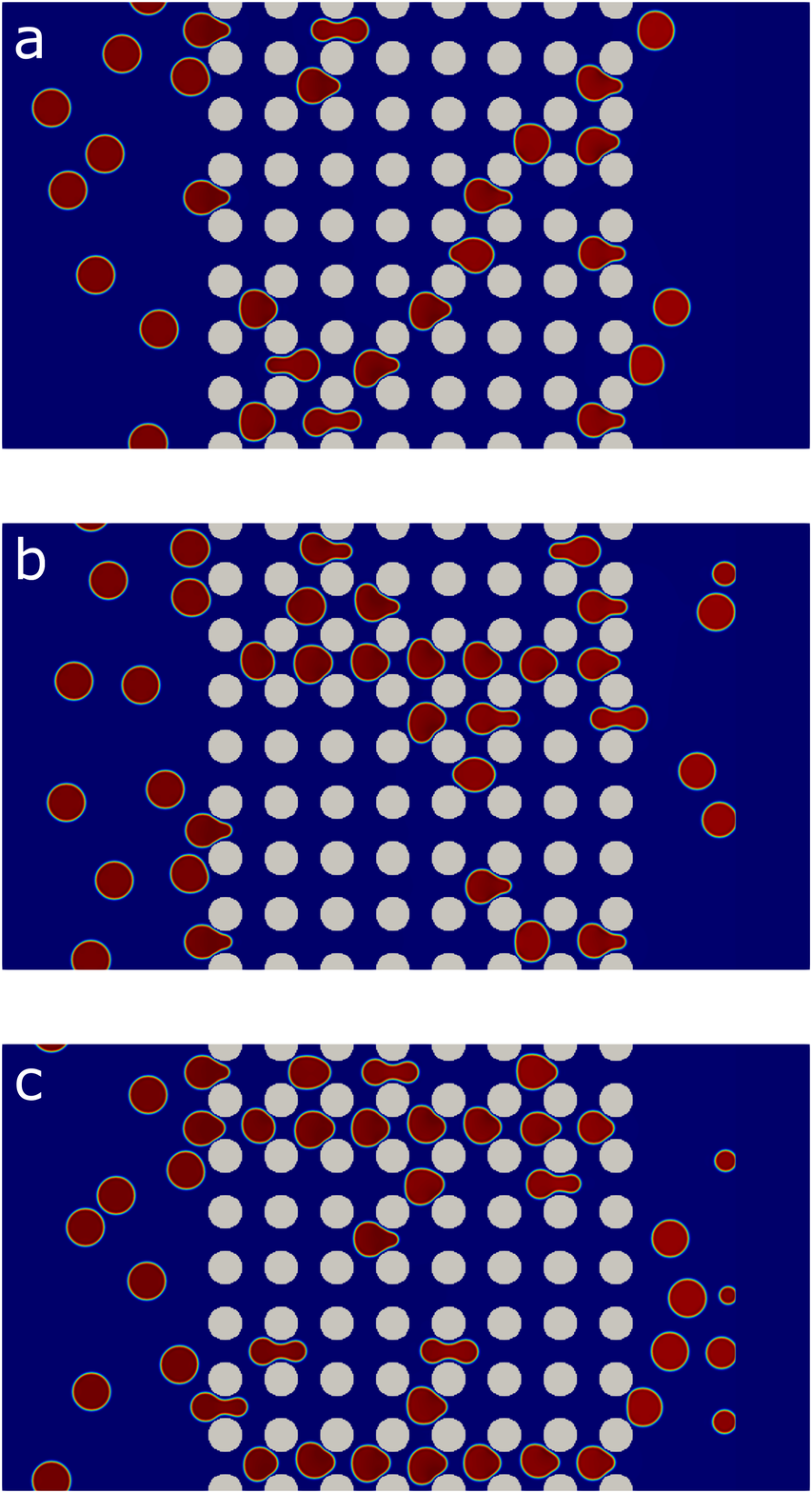}
	\caption{Snapshots of the simulation with $v_b=0.001$ at three different times. a) $t=345000$, collective flow of droplets in the diagonal direction. b) $t=1915000$, collective flow of one  line of droplets in the direction of the flow. c) $t=4425000$, collective flow of two lines of droplets.}\label{fig-examples}
\end{figure}

Now, we analyse the behaviour of many droplets. The frequency of droplet addition depends on the boundary velocity in order to keep the average number of droplets constant in the medium in the absence of obstacles. In order to do that, we consider a velocity-dependent frequency: $f=v_b/10$.
In Fig.~\ref{fig-vel-num}a, we plot the average velocity of the droplets $v_d$ in the steady state as a function of the boundary velocity for three different frequencies. The averages are in time and space. Note that at low velocities (below $v_b=0.0017$) one single droplet is trapped while many droplets can flow through the obstacles. This implies that the droplets behave collectively and interact through the fluid velocity. Above this threshold the average droplet velocity increases linearly with the boundary velocity, in line with the velocity of a single droplet. The effect of the frequency on the average velocity is small, and the velocity increases only slightly with the frequency. In addition, the velocity of many droplets is always higher than that of a single droplet. This occurs because the fluid velocity at the free pores increases when some of them are blocked by the droplets. At higher frequencies, there are more droplets blocking the pores making the average velocity to increase.
Fig.~\ref{fig-hist} displays histograms of the droplet velocities for two boundary velocities. It is clear that most of the droplets are at rest or have a velocity lower than that of the boundary.
The average number of droplets in the porous medium (the region with obstacles) is calculated for the three frequencies and is plotted in Fig.~\ref{fig-vel-num}b. This number decreases as the boundary velocity increases and the effect is larger at low velocities. This may be understood as the droplets cluster when they cooperate, i.e., at low velocities. The clustering also occurs in the form of static horizontal lines of droplets as shown in Fig.~\ref{fig-examples}.

As an example, Fig.~\ref{fig-examples} shows snapshots of the simulations with $v_b=0.001$ obtained at three different times. For this value of the velocity, one single droplet always gets trapped by the first layer of obstacles. However, as shown in the snapshots, with several droplets, droplets are only trapped for a finite time. The cooperative flow of the droplets may occur through their alignment in the vertical or diagonal directions as shown in Fig.~\ref{fig-examples}a. When one of the droplets is trapped by the obstacles (usually, the last droplet of the diagonal array of droplets is trapped while the others flow), incoming droplets will align behind it forming a horizontal line (Fig.~\ref{fig-examples}b). These lines remain static for a certain time, which is $\sim 2\times10^6$ iterations in the example of Fig.~\ref{fig-examples}, and may be destroyed and formed at different heights (Fig.~\ref{fig-examples}c). These lines occur for velocities below a certain threshold. At higher velocities, the individual droplets can flow through the obstacles and they interact weakly.

\begin{figure}
	\centering
	\includegraphics[width=\columnwidth]{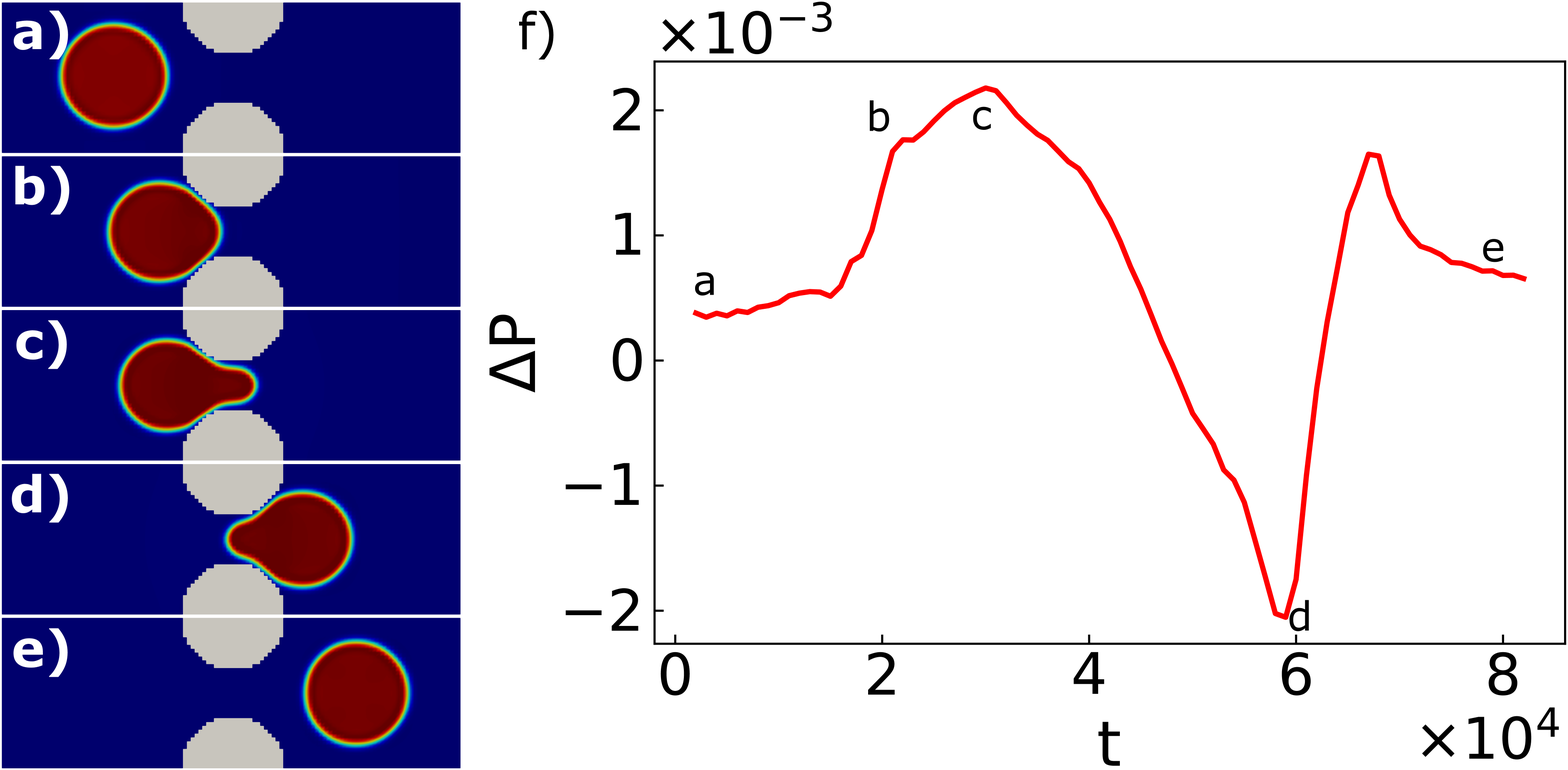}
	\caption{Snapshots of one droplet flowing through a single pore at a) $t=1000$, b) $t=20000$, c) $t=30000$, d) $t=60000$, e) $t=80000$. f) Time evolution of the pressure difference before and after the droplet flows through the pore.}
	\label{fig-dp-1d}
\end{figure}

We investigate the mechanism that drives the cooperation of droplets flowing through obstacles. We start by simulating the flow of one droplet though a single pore with constant velocity $v_b=0.0005$ at the inlet and the outlet and periodic boundary conditions on the top and bottom. The system size is $L_X \times L_Y = 120 \times 40$, the circular obstacles, with $R_o=13$, are separated by $D=14$ and the radius of the droplet is $R_d=14$. Fig.~\ref{fig-dp-1d} depicts snapshots at different times and the pressure difference across the length of the pore (on the top) as a function of time. As the droplet is larger than the pore size, it has to deform as it flows through and the pressure difference has to compensate the resistance to deformation quantified by the surface tension. The results in Fig.~\ref{fig-dp-1d}f show that, after trapping the droplet ($\sim t=20000$), the pressure difference increases until the droplet reaches the neck of the pore. Then, the pressure difference drops to a minimum as the droplet exits the neck and it returns to the initial value after the droplet flows away. 

\begin{figure}
	\centering
	\includegraphics[width=\columnwidth]{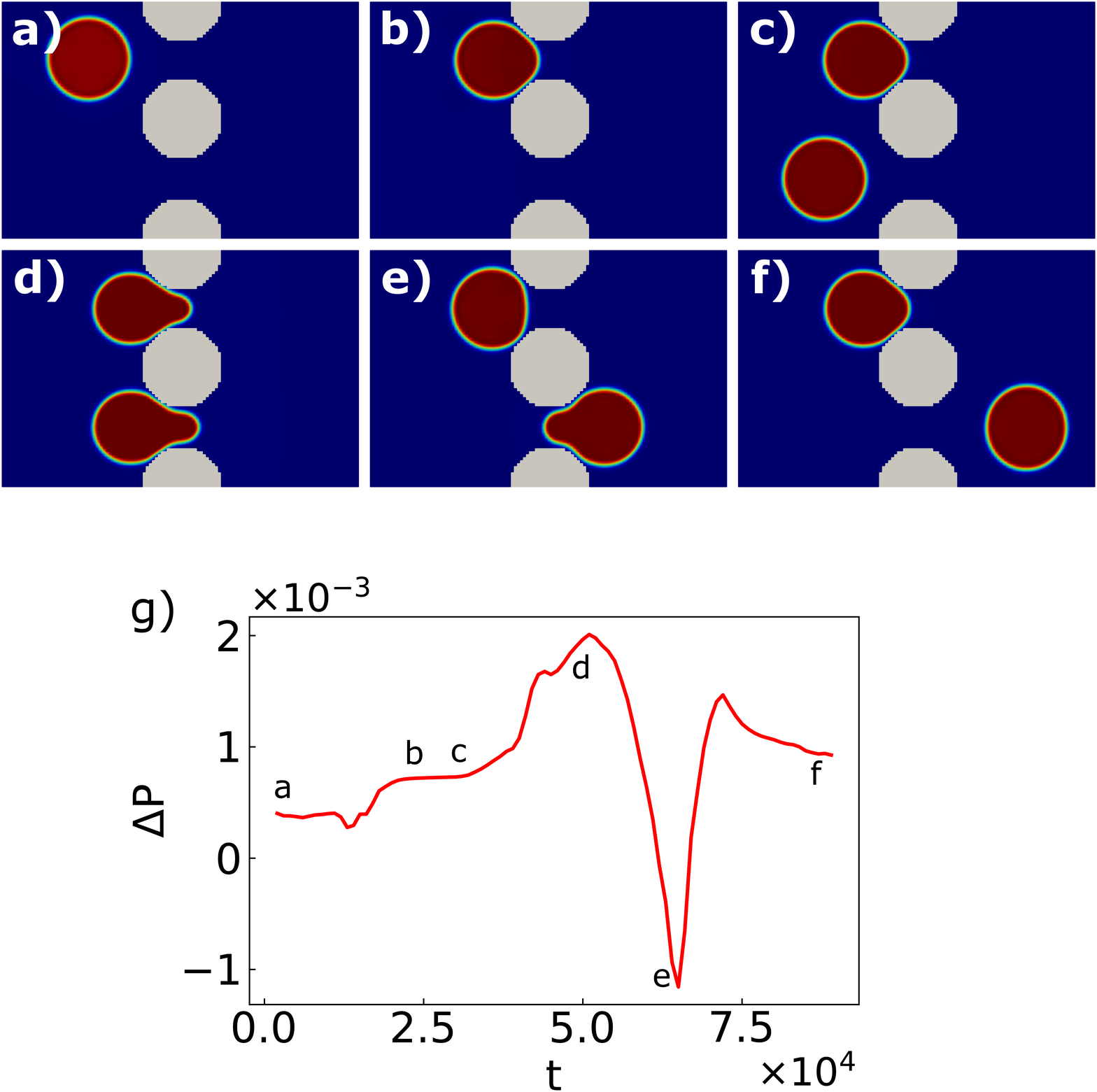}
	\caption{Snapshots of the droplets flowing through the pores: a) $t=1000$, the first droplet is created; b) $t=25000$, the first droplet is trapped; c) $t=31000$, the second droplet is created; d) $t=50000$, the two droplets are trapped; e) $t=65000$, one of the droplets flows when the pressure difference reaches the threshold; f) $t=85000$. g) Time evolution of the pressure difference  before and after the droplet flows through the pore.}\label{fig-dp-2d}
\end{figure}

Now we consider two pores as illustrated in Fig.~\ref{fig-dp-2d}. We start with one droplet at the top with $v_b=0.0005$ in a domain $L_X\times L_Y = 120\times 80$. At short times, the droplet is trapped as before while the fluid flows through the second pore. The results plotted in Fig.~\ref{fig-dp-2d}g reveal that the pressure difference remains constant between $t=20000$ and $t=30000$ as the fluid flow through the second pore does not allow it to increase. At $t=30000$, a second droplet is created at the bottom and it is trapped by the open pore, blocking the fluid flow. This increases the pressure difference across the pores until one of the droplets (the second one in this case) flows through, after which the pressure difference returns to its initial value. As discussed in the previous section, when droplets are trapped blocking the fluid flow, the local pressure difference increases, allowing nearby droplets to deform and flow through the porous medium.

\begin{figure}
	\centering
	\includegraphics[width=\columnwidth]{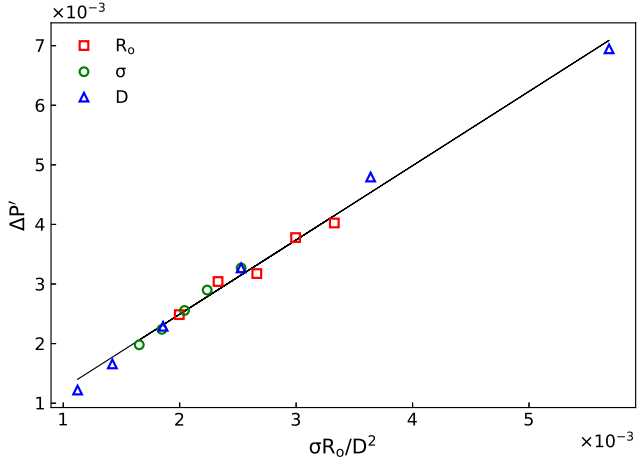}
	\caption{Calculation of the minimum Bond number required for the flow of one droplet. The threshold pressure difference $\Delta P^\prime$ was calculated as a function of each of the different parameters (symbols) while keeping the others fixed. }\label{fig-Bo-number}
\end{figure}

Note that the pressure difference required for the flow of one droplet $\Delta P^\prime$ is roughly the same in both cases. We can write the Bond number as the ratio of the external driving force due to the pressure field and the the resisting surface tension. There are two relevant length scales in the system considered here: the size of the pore neck, which is set by $D$ and the length of the pore over which the pressure field varies, which is set by $R_o$. By replacing $g\Delta \rho $ with $\Delta P/R_o$ in the gravitational Bond number~\cite{Coelho_2022}, we obtain:
\begin{equation}
    \rm{Bo} = \frac{D^2\Delta P}{R_o \sigma }.
    \label{bond-eq}
\end{equation}
This may be used to obtain the threshold pressure difference $\Delta P^\prime$ required for droplet flow through the pores in terms of the other parameters. We calculate the threshold $\rm{Bo}$ required for droplet flow, through simulations of a single droplet, by varying each one of the parameters ($D$, $R_o$ and $\sigma$) while keeping the others fixed. We used the following range of parameters: $12 < R_o < 20$, $0.017 < \sigma < 0.026$ and $8<D<18$. The results for the threshold $\Delta P^\prime$ as a function of $\sigma R_o/D^2$ collapse into a straight line as shown in Fig.~\ref{fig-Bo-number}. From a linear fit, we obtain $\rm{Bo} \approx 1.25 \pm 0.01$ at threshold. Thus, droplets will flow for pressure differences larger than:
\begin{equation}
    \Delta P^\prime = 1.25 \frac{\sigma R_o}{D^2}.
    \label{pressure-prime-eq}
\end{equation}

\begin{figure}
	\centering
	\includegraphics[width=\columnwidth]{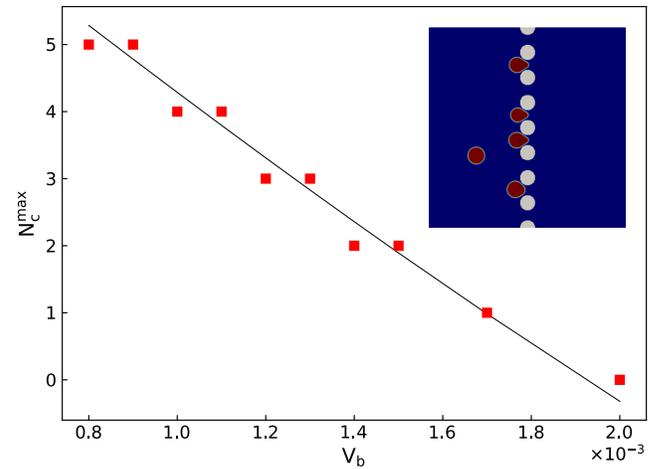}
	\caption{Maximum number of clogged pores as a function of the boundary velocity for a column of obstacles. The solid line is a  fit of Eq.~\eqref{pressure-pm-eq} with the correction factor $C=146 \pm 4$. Inset: snapshot of the simulation for the velocity $v_b=0.001$. }\label{fig-ncmax}
\end{figure}

We will now obtain a similar relation for the flow of many droplets in a lattice of obstacles.
Consider that the average velocity in a given pore is $v_p$. The simplest relation between $v_p$ and the pressure difference over the pore is given by the Bernoulli equation. This is a crude approximation since the fluid is viscous and a phase separated binary mixture (note that the pressure is given by Eq.~\eqref{pressure-eq}). We proceed by multiplying the Bernoulli pressure difference by a correction factor $C$ to be calculated later: 
\begin{equation*}
    \Delta P \approx \frac{C \rho}{2} (v_p^2-v_b^2).
\end{equation*}
Mass conservation implies that the flow rate at the inlet does not vary on any plane perpendicular to the flow. Thus, on a plane containing an array of obstacles:
\begin{equation*}
    v_b L_Y = v_p \left( L_Y - 2 N_o R_o - N_c D \right),
\end{equation*}
where $N_o$ is the number of obstacles in the perpendicular direction to the flow and $N_c$ is the number of clogged pores (with trapped droplets). Then:
\begin{equation}
\Delta P = \frac{C\rho v_b^2}{2\left(1-2 \frac{N_o}{L_Y}R_o - \frac{N_c}{L_Y}D \right )^2} - \frac{C\rho v_b^2}{2}
\label{pressure-pm-eq}
\end{equation}
Note that $N_o/L_Y$ and $N_c/L_Y$ may be interpreted as the number of obstacles and of clogged pores per unit length in any porous medium. One concludes that when the boundary velocity $v_b$ is not sufficient to reach the threshold $\Delta P^\prime$, then $N_c/L_Y$ has to increase until that condition is met. To obtain the correction factor $C$, we simulate one column of obstacles and create droplets one by one until one of them flows through. The maximum number of clogged pores $N_c^{\text{max}}$ is then related to $v_b$ and a fit of Eq.~\eqref{pressure-pm-eq} used to obtain $C$ as shown in Fig.~\ref{fig-ncmax}. We obtain $C=146 \pm 4$. We assumed that the droplets cooperate by forming a vertical line of droplets but, as we have shown, they can also form diagonal lines (Fig.~\ref{fig-examples}). Eq.~\ref{pressure-pm-eq} gives an estimate of the number of droplets required to cooperate (i.e., to clog pores) to flow in a lattice of obstacles, as a function of the boundary velocity. The case of horizontal lines in the porous medium would be represented here by one single clogged pore.

\begin{figure}
	\centering
	\includegraphics[width=\columnwidth]{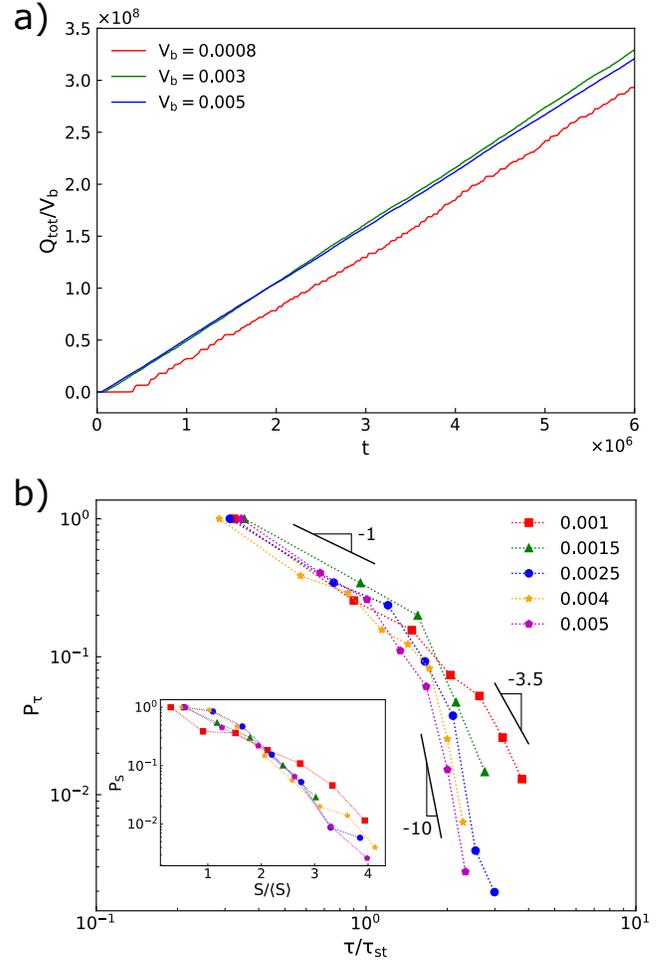}
	\caption{a) Total volume of droplets that flow through the porous medium divided by the boundary velocity versus time. b) Cumulative distribution function of the time lapses between the avalanches at the end of the obstacles ($x=400$) for five inlet velocities (legend). The time is given in Stokes time, $\tau_{st} =R_o/v_b$, the typical time taken by a droplet of radius R to move a distance equal to its own size. The solid lines indicate the slopes of three power laws for comparison. The inset shows the histogram of the avalanches areas $S$ rescaled by the average avalanche area $\langle S\rangle$.}\label{fig-avalanche}
\end{figure}
Additionally, we investigated the occurrence of bursts in the droplet flow. Bursts occur when, for instance, many particles pass through a constriction~\cite{Zuriguel2014}. Here, as the particles are flowing droplets, it is natural to use the instantaneous flow rate through a given plane, $Q(t)=\int \rho v_d dS$, to analyze the avalanches (passage of one or more droplets together). Figure~\ref{fig-avalanche}a shows the total volume of droplets that flows though the porous medium, $Q_{\text{tot}}=\int Q(t) dt$, as a function of time for three different boundary velocities. We divide $Q_{\text{tot}}$ by the boundary velocity to compare the curves. The curves with $v_b>0.0017$ (threshold below which the droplets cooperate) are approximately linear and  collapse as $Q_{\text{tot}}$ becomes proportional to $v_b$. The curve for $v_b=0.0008$ (below threshold) is composed by steps with short time intervals for which $Q(t)$ is zero (no droplet flows). 
There are two quantities of interest: the time lapses between subsequent avalanches (time between subsequent zero flow rates) and the area of the avalanches (integral over time of the flow rate between subsequent zeros). If there are bursts, the cumulative distribution function (CDF) of the avalanche areas is exponential and the CDF of the time lapses exhibits a power law tail. We analyzed the flow rate of the droplets at the end of the lattice of obstacles (at $x=400$). The CDF of the time lapses $\tau$ and the area of the avalanches are shown in Fig.~\ref{fig-avalanche}. The results suggest that there are no bursts of droplets flowing through a lattice of obstacles. This might be due to the constant flow rate used in the boundary conditions. With constant a flow rate, the pressure increases to guarantee that the flow rate is constant. With constant pressure drop, large fluctuations in the flow are expected with the possibility of bursts~\cite{PhysRevLett.117.275702}.

\section{Conclusion}
\label{conclusion.sec}

We addressed the cooperation of immiscible droplets flowing through porous media under conditions for which a single droplet is trapped. At high boundary velocities, the droplets flow almost independently with little evidence of cooperative or collective behavior. At low velocities, however, the droplets block a number of pores and as a result increase the flow velocity in the others. This occurs until the pressure difference over the pores is large enough to allow one droplet to deform and then flow through. The result clearly illustrates the hydrodynamic interaction between droplets, which are not at close contact. In this regime, the droplets cooperate by blocking some of the pores and thus promote the conditions for droplet deformation and flow. 

The conditions for a single droplet to flow through one pore were analyzed, considering the geometrical and the fluid properties. We found a modified Bond number that takes into account the pressure field (instead of the gravitational one) as the relevant non-dimensional number to characterize droplet flow, for a wide range of parameters. Moreover, the mechanism of cooperation was investigated and a general expression for the threshold pressure difference for droplet flow in porous media was obtained. 

The 2D simulations reported here model 3D droplets flowing though pores with aspect ratio (height/width, in the directions perpendicular to the flow) close to one. As pointed out in Ref.~\cite{C9SM00300B}, if the aspect ratio is larger than one, two droplets cooperate to flow through the same pore, which may be used for size sorting of deformable particles. The latter study considered a single pore, rendering the application of the results difficult in real porous media with multiple pores. In fact it is not obvious that in 3D the droplets or flexible particles exhibit similar behavior and thus limiting the size sorting application.  

This work paves the way for future studies involving multicomponent flows in porous media in order to enhance oil extraction. One next step in this direction would be to simulate fluid displacement together with the droplets made of another component. The pore sizes should also be heterogeneous in order to mimic the realistic conditions of a porous medium. 

Another important application of flowing deformable particles in porous media is
the sorting of cells in microfluidic devices. Circulating cells in the human blood system are strongly deformed during each passage across the microvasculature of the organs. However, pathologies such as pneumonia and cancer, are characterized by stiffer cells that are trapped in the microvasculature triggering serious complications~\cite{Lam2008, Rosenbluth2008, Wong2002, doi:10.1504/IJNT.2012.045340, Hogg1987, Hotchkiss2003, Yoshida2006}. Thus, the separation of cells by their stiffness is highly relevant in novel clinical investigation. Preira et al. proposed a microfluidic method to sort the cells of non-adherent cell populations by deformability~\cite{C2LC40847C}. It consists of a porous medium with a gradient in pore sizes so that stiffer cells are trapped in the region of narrower pores. The results reported here can be extended to cell sorting since the passage of the cells may also depend on collective behaviour in dense suspensions.

\section{Appendix}

Tables~\ref{d3q41-table} and~\ref{d3q39-table} provide the velocity vectors and weights of the two lattices used in the LBM simulations. 

\begin{table}[h!]
\caption{Velocity vectors and weights for the D3Q41 lattice. The speed of sound $c_{s}^2$ is $1-\sqrt{2/5}$.}
\begin{tabular} {|c|c|}
\hline
 $\mathbf{c}_{i}$ & $w_{i}$ \\ \hline
 $(0,0,0)$ & $2(5045-1507 \sqrt{10}) / 2025$ \\
 $(\pm 1,0,0),(0,\pm 1,0),(0,0,\pm 1)$  & $377 /(5 \sqrt{10})-(91 / 40)$ \\
 $(\pm 1,\pm 1,0),(\pm 1,0,\pm 1),(0,\pm 1,\pm 1)$ & $(55-17 \sqrt{10}) / 50$ \\
 $(\pm 1,\pm 1,\pm 1)$ & (233 $\sqrt{10}-730) / 1600$ \\
 $(\pm 3,0,0),(0,\pm 3,0),(0,0,\pm 3)$  & $(295-92 \sqrt{10}) / 16200$ \\ 
 $(\pm 3,\pm 3,\pm 3)$  & $(130-41 \sqrt{10}) / 129600$ \\ \hline
\end{tabular}
\label{d3q41-table}
\end{table}

\begin{table}[h!]
\caption{Velocity vectors and weights for the D3Q39 lattice. The speed of sound $c_{s}^2$ is $2/3$.}
\begin{tabular} {|c|c|}
\hline
$\mathbf{c}_{i}$ & $w_{i}$\\ \hline
 $(0,0,0)$ & 1/12 \\ 
 $(\pm 1,0,0),(0,\pm 1,0),(0,0,\pm 1)$  & 1/12 \\
 $(\pm 1,\pm 1,\pm 1)$ & 1/27 \\
 $(\pm 2,0,0),(0,\pm 2,0),(0,0,\pm 2)$  & 2/135 \\
 $(\pm 2,\pm 2,0),(\pm 2,0,\pm 2),(0,\pm 2,\pm 2)$ & 1/142 \\
 $(\pm 3,0,0),(0,\pm 3,0),(0,0,\pm 3)$  & 1/1620 \\ \hline
\end{tabular}
\label{d3q39-table}
\end{table}

\section*{Acknowledgements}

We acknowledge financial support from the Portuguese Foundation for Science and Technology (FCT) under the contracts: EXPL/FIS-MAC/0406/2021, PTDC/FIS-MAC/28146/2017 (LISBOA-01-0145-FEDER-028146), PTDC/FISMAC/5689/2020, 2020.08525.BD, UIDB/00618/2020 and UIDP/00618/2020.

\section*{Data availability}

The data that support the findings of this study are available from the corresponding author upon reasonable request.

\nocite{*}
\bibliography{aipsamp}% Produces the bibliography via BibTeX.

%aipnum4-2.bst 2019-01-14 (MD) hand-edited version of apsrev4-1.bst
\begin{thebibliography}{37}%
\makeatletter
\providecommand \@ifxundefined [1]{%
 \@ifx{#1\undefined}
}%
\providecommand \@ifnum [1]{%
 \ifnum #1\expandafter \@firstoftwo
 \else \expandafter \@secondoftwo
 \fi
}%
\providecommand \@ifx [1]{%
 \ifx #1\expandafter \@firstoftwo
 \else \expandafter \@secondoftwo
 \fi
}%
\providecommand \natexlab [1]{#1}%
\providecommand \enquote  [1]{``#1''}%
\providecommand \bibnamefont  [1]{#1}%
\providecommand \bibfnamefont [1]{#1}%
\providecommand \citenamefont [1]{#1}%
\providecommand \href@noop [0]{\@secondoftwo}%
\providecommand \href [0]{\begingroup \@sanitize@url \@href}%
\providecommand \@href[1]{\@@startlink{#1}\@@href}%
\providecommand \@@href[1]{\endgroup#1\@@endlink}%
\providecommand \@sanitize@url [0]{\catcode `\\12\catcode `\$12\catcode
  `\&12\catcode `\#12\catcode `\^12\catcode `\_12\catcode `\%12\relax}%
\providecommand \@@startlink[1]{}%
\providecommand \@@endlink[0]{}%
\providecommand \url  [0]{\begingroup\@sanitize@url \@url }%
\providecommand \@url [1]{\endgroup\@href {#1}{\urlprefix }}%
\providecommand \urlprefix  [0]{URL }%
\providecommand \Eprint [0]{\href }%
\providecommand \doibase [0]{https://doi.org/}%
\providecommand \selectlanguage [0]{\@gobble}%
\providecommand \bibinfo  [0]{\@secondoftwo}%
\providecommand \bibfield  [0]{\@secondoftwo}%
\providecommand \translation [1]{[#1]}%
\providecommand \BibitemOpen [0]{}%
\providecommand \bibitemStop [0]{}%
\providecommand \bibitemNoStop [0]{.\EOS\space}%
\providecommand \EOS [0]{\spacefactor3000\relax}%
\providecommand \BibitemShut  [1]{\csname bibitem#1\endcsname}%
\let\auto@bib@innerbib\@empty
%</preamble>
\bibitem [{\citenamefont {Kamal}\ \emph {et~al.}(2015)\citenamefont {Kamal},
  \citenamefont {Sultan}, \citenamefont {Al-Mubaiyedh},\ and\ \citenamefont
  {Hussein}}]{Kamal2015}%
  \BibitemOpen
  \bibfield  {author} {\bibinfo {author} {\bibfnamefont {M.~S.}\ \bibnamefont
  {Kamal}}, \bibinfo {author} {\bibfnamefont {A.~S.}\ \bibnamefont {Sultan}},
  \bibinfo {author} {\bibfnamefont {U.~A.}\ \bibnamefont {Al-Mubaiyedh}},\ and\
  \bibinfo {author} {\bibfnamefont {I.~A.}\ \bibnamefont {Hussein}},\
  }\bibfield  {title} {\enquote {\bibinfo {title} {Review on polymer flooding:
  Rheology, adsorption, stability, and field applications of various polymer
  systems},}\ }\href {https://doi.org/10.1080/15583724.2014.982821} {\bibfield
  {journal} {\bibinfo  {journal} {Polymer Reviews}\ }\textbf {\bibinfo {volume}
  {55}},\ \bibinfo {pages} {491--530} (\bibinfo {year} {2015})}\BibitemShut
  {NoStop}%
\bibitem [{\citenamefont {Xia}\ \emph {et~al.}(2008)\citenamefont {Xia},
  \citenamefont {Wang}, \citenamefont {Wang}, \citenamefont {guo Ma},
  \citenamefont {Deng},\ and\ \citenamefont {Liu}}]{Xia2008}%
  \BibitemOpen
  \bibfield  {author} {\bibinfo {author} {\bibfnamefont {H.}~\bibnamefont
  {Xia}}, \bibinfo {author} {\bibfnamefont {D.}~\bibnamefont {Wang}}, \bibinfo
  {author} {\bibfnamefont {G.}~\bibnamefont {Wang}}, \bibinfo {author}
  {\bibfnamefont {W.}~\bibnamefont {guo Ma}}, \bibinfo {author} {\bibfnamefont
  {H.~W.}\ \bibnamefont {Deng}},\ and\ \bibinfo {author} {\bibfnamefont
  {J.}~\bibnamefont {Liu}},\ }\bibfield  {title} {\enquote {\bibinfo {title}
  {Mechanism of the effect of micro-forces on residual oil in chemical
  flooding},}\ }in\ \href {https://doi.org/10.2118/114335-ms} {\emph {\bibinfo
  {booktitle} {All Days}}}\ (\bibinfo  {publisher} {{SPE}},\ \bibinfo {year}
  {2008})\BibitemShut {NoStop}%
\bibitem [{\citenamefont {Wever}, \citenamefont {Picchioni},\ and\
  \citenamefont {Broekhuis}(2011)}]{WEVER20111558}%
  \BibitemOpen
  \bibfield  {author} {\bibinfo {author} {\bibfnamefont {D.}~\bibnamefont
  {Wever}}, \bibinfo {author} {\bibfnamefont {F.}~\bibnamefont {Picchioni}},\
  and\ \bibinfo {author} {\bibfnamefont {A.}~\bibnamefont {Broekhuis}},\
  }\bibfield  {title} {\enquote {\bibinfo {title} {Polymers for enhanced oil
  recovery: A paradigm for structure–property relationship in aqueous
  solution},}\ }\href
  {https://doi.org/https://doi.org/10.1016/j.progpolymsci.2011.05.006}
  {\bibfield  {journal} {\bibinfo  {journal} {Progress in Polymer Science}\
  }\textbf {\bibinfo {volume} {36}},\ \bibinfo {pages} {1558--1628} (\bibinfo
  {year} {2011})},\ \bibinfo {note} {special Topic: Energy Related
  Materials}\BibitemShut {NoStop}%
\bibitem [{\citenamefont {Han}\ \emph {et~al.}(1999)\citenamefont {Han},
  \citenamefont {Yang}, \citenamefont {Zhang}, \citenamefont {Lou},\ and\
  \citenamefont {Chang}}]{Han1999}%
  \BibitemOpen
  \bibfield  {author} {\bibinfo {author} {\bibfnamefont {D.-K.}\ \bibnamefont
  {Han}}, \bibinfo {author} {\bibfnamefont {C.-Z.}\ \bibnamefont {Yang}},
  \bibinfo {author} {\bibfnamefont {Z.-Q.}\ \bibnamefont {Zhang}}, \bibinfo
  {author} {\bibfnamefont {Z.-H.}\ \bibnamefont {Lou}},\ and\ \bibinfo {author}
  {\bibfnamefont {Y.-I.}\ \bibnamefont {Chang}},\ }\bibfield  {title} {\enquote
  {\bibinfo {title} {Recent development of enhanced oil recovery in china},}\
  }\href {https://doi.org/10.1016/s0920-4105(98)00067-9} {\bibfield  {journal}
  {\bibinfo  {journal} {Journal of Petroleum Science and Engineering}\ }\textbf
  {\bibinfo {volume} {22}},\ \bibinfo {pages} {181--188} (\bibinfo {year}
  {1999})}\BibitemShut {NoStop}%
\bibitem [{\citenamefont {Browne}, \citenamefont {Shih},\ and\ \citenamefont
  {Datta}(2020{\natexlab{a}})}]{browne_shih_datta_2020}%
  \BibitemOpen
  \bibfield  {author} {\bibinfo {author} {\bibfnamefont {C.~A.}\ \bibnamefont
  {Browne}}, \bibinfo {author} {\bibfnamefont {A.}~\bibnamefont {Shih}},\ and\
  \bibinfo {author} {\bibfnamefont {S.~S.}\ \bibnamefont {Datta}},\ }\bibfield
  {title} {\enquote {\bibinfo {title} {Bistability in the unstable flow of
  polymer solutions through pore constriction arrays},}\ }\href
  {https://doi.org/10.1017/jfm.2020.122} {\bibfield  {journal} {\bibinfo
  {journal} {Journal of Fluid Mechanics}\ }\textbf {\bibinfo {volume} {890}},\
  \bibinfo {pages} {A2} (\bibinfo {year} {2020}{\natexlab{a}})}\BibitemShut
  {NoStop}%
\bibitem [{\citenamefont {Browne}, \citenamefont {Shih},\ and\ \citenamefont
  {Datta}(2020{\natexlab{b}})}]{https://doi.org/10.1002/smll.201903944}%
  \BibitemOpen
  \bibfield  {author} {\bibinfo {author} {\bibfnamefont {C.~A.}\ \bibnamefont
  {Browne}}, \bibinfo {author} {\bibfnamefont {A.}~\bibnamefont {Shih}},\ and\
  \bibinfo {author} {\bibfnamefont {S.~S.}\ \bibnamefont {Datta}},\ }\bibfield
  {title} {\enquote {\bibinfo {title} {Pore-scale flow characterization of
  polymer solutions in microfluidic porous media},}\ }\href
  {https://doi.org/https://doi.org/10.1002/smll.201903944} {\bibfield
  {journal} {\bibinfo  {journal} {Small}\ }\textbf {\bibinfo {volume} {16}},\
  \bibinfo {pages} {1903944} (\bibinfo {year}
  {2020}{\natexlab{b}})}\BibitemShut {NoStop}%
\bibitem [{\citenamefont {Chakrabarti}, \citenamefont {Gaillard},\ and\
  \citenamefont {Saintillan}(2020)}]{D0SM00390E}%
  \BibitemOpen
  \bibfield  {author} {\bibinfo {author} {\bibfnamefont {B.}~\bibnamefont
  {Chakrabarti}}, \bibinfo {author} {\bibfnamefont {C.}~\bibnamefont
  {Gaillard}},\ and\ \bibinfo {author} {\bibfnamefont {D.}~\bibnamefont
  {Saintillan}},\ }\bibfield  {title} {\enquote {\bibinfo {title} {Trapping{,}
  gliding{,} vaulting: transport of semiflexible polymers in periodic post
  arrays},}\ }\href {https://doi.org/10.1039/D0SM00390E} {\bibfield  {journal}
  {\bibinfo  {journal} {Soft Matter}\ }\textbf {\bibinfo {volume} {16}},\
  \bibinfo {pages} {5534--5544} (\bibinfo {year} {2020})}\BibitemShut {NoStop}%
\bibitem [{\citenamefont {Lei}\ \emph {et~al.}(2020)\citenamefont {Lei},
  \citenamefont {Liu}, \citenamefont {Xie}, \citenamefont {Yang}, \citenamefont
  {Wu},\ and\ \citenamefont {Wang}}]{https://doi.org/10.1002/ese3.563}%
  \BibitemOpen
  \bibfield  {author} {\bibinfo {author} {\bibfnamefont {W.}~\bibnamefont
  {Lei}}, \bibinfo {author} {\bibfnamefont {T.}~\bibnamefont {Liu}}, \bibinfo
  {author} {\bibfnamefont {C.}~\bibnamefont {Xie}}, \bibinfo {author}
  {\bibfnamefont {H.}~\bibnamefont {Yang}}, \bibinfo {author} {\bibfnamefont
  {T.}~\bibnamefont {Wu}},\ and\ \bibinfo {author} {\bibfnamefont
  {M.}~\bibnamefont {Wang}},\ }\bibfield  {title} {\enquote {\bibinfo {title}
  {Enhanced oil recovery mechanism and recovery performance of micro-gel
  particle suspensions by microfluidic experiments},}\ }\href
  {https://doi.org/https://doi.org/10.1002/ese3.563} {\bibfield  {journal}
  {\bibinfo  {journal} {Energy Science \& Engineering}\ }\textbf {\bibinfo
  {volume} {8}},\ \bibinfo {pages} {986--998} (\bibinfo {year}
  {2020})}\BibitemShut {NoStop}%
\bibitem [{\citenamefont {Bai}\ \emph {et~al.}(2007{\natexlab{a}})\citenamefont
  {Bai}, \citenamefont {Liu}, \citenamefont {Coste},\ and\ \citenamefont
  {Li}}]{Bai2007}%
  \BibitemOpen
  \bibfield  {author} {\bibinfo {author} {\bibfnamefont {B.}~\bibnamefont
  {Bai}}, \bibinfo {author} {\bibfnamefont {Y.}~\bibnamefont {Liu}}, \bibinfo
  {author} {\bibfnamefont {J.~P.}\ \bibnamefont {Coste}},\ and\ \bibinfo
  {author} {\bibfnamefont {L.}~\bibnamefont {Li}},\ }\bibfield  {title}
  {\enquote {\bibinfo {title} {Preformed particle gel for conformance control:
  Transport mechanism through porous media},}\ }\href
  {https://doi.org/10.2118/89468-pa} {\bibfield  {journal} {\bibinfo  {journal}
  {{SPE} Reservoir Evaluation {\&} Engineering}\ }\textbf {\bibinfo {volume}
  {10}},\ \bibinfo {pages} {176--184} (\bibinfo {year}
  {2007}{\natexlab{a}})}\BibitemShut {NoStop}%
\bibitem [{\citenamefont {Bai}\ \emph {et~al.}(2007{\natexlab{b}})\citenamefont
  {Bai}, \citenamefont {Li}, \citenamefont {Liu}, \citenamefont {Liu},
  \citenamefont {Wang},\ and\ \citenamefont {You}}]{Bai2007-2}%
  \BibitemOpen
  \bibfield  {author} {\bibinfo {author} {\bibfnamefont {B.}~\bibnamefont
  {Bai}}, \bibinfo {author} {\bibfnamefont {L.}~\bibnamefont {Li}}, \bibinfo
  {author} {\bibfnamefont {Y.}~\bibnamefont {Liu}}, \bibinfo {author}
  {\bibfnamefont {H.}~\bibnamefont {Liu}}, \bibinfo {author} {\bibfnamefont
  {Z.}~\bibnamefont {Wang}},\ and\ \bibinfo {author} {\bibfnamefont
  {C.}~\bibnamefont {You}},\ }\bibfield  {title} {\enquote {\bibinfo {title}
  {Preformed particle gel for conformance control: Factors affecting its
  properties and applications},}\ }\href {https://doi.org/10.2118/89389-pa}
  {\bibfield  {journal} {\bibinfo  {journal} {{SPE} Reservoir Evaluation {\&}
  Engineering}\ }\textbf {\bibinfo {volume} {10}},\ \bibinfo {pages} {415--422}
  (\bibinfo {year} {2007}{\natexlab{b}})}\BibitemShut {NoStop}%
\bibitem [{\citenamefont {Li}\ \emph {et~al.}(2020)\citenamefont {Li},
  \citenamefont {Yu}, \citenamefont {Li}, \citenamefont {Chen}, \citenamefont
  {Deng}, \citenamefont {Anbari},\ and\ \citenamefont
  {Fan}}]{doi:10.1063/1.5139887}%
  \BibitemOpen
  \bibfield  {author} {\bibinfo {author} {\bibfnamefont {S.}~\bibnamefont
  {Li}}, \bibinfo {author} {\bibfnamefont {H.}~\bibnamefont {Yu}}, \bibinfo
  {author} {\bibfnamefont {T.-D.}\ \bibnamefont {Li}}, \bibinfo {author}
  {\bibfnamefont {Z.}~\bibnamefont {Chen}}, \bibinfo {author} {\bibfnamefont
  {W.}~\bibnamefont {Deng}}, \bibinfo {author} {\bibfnamefont {A.}~\bibnamefont
  {Anbari}},\ and\ \bibinfo {author} {\bibfnamefont {J.}~\bibnamefont {Fan}},\
  }\bibfield  {title} {\enquote {\bibinfo {title} {Understanding transport of
  an elastic, spherical particle through a confining channel},}\ }\href
  {https://doi.org/10.1063/1.5139887} {\bibfield  {journal} {\bibinfo
  {journal} {Applied Physics Letters}\ }\textbf {\bibinfo {volume} {116}},\
  \bibinfo {pages} {103705} (\bibinfo {year} {2020})}\BibitemShut {NoStop}%
\bibitem [{\citenamefont {Perazzo}\ \emph {et~al.}(2018)\citenamefont
  {Perazzo}, \citenamefont {Tomaiuolo}, \citenamefont {Preziosi},\ and\
  \citenamefont {Guido}}]{PERAZZO2018305}%
  \BibitemOpen
  \bibfield  {author} {\bibinfo {author} {\bibfnamefont {A.}~\bibnamefont
  {Perazzo}}, \bibinfo {author} {\bibfnamefont {G.}~\bibnamefont {Tomaiuolo}},
  \bibinfo {author} {\bibfnamefont {V.}~\bibnamefont {Preziosi}},\ and\
  \bibinfo {author} {\bibfnamefont {S.}~\bibnamefont {Guido}},\ }\bibfield
  {title} {\enquote {\bibinfo {title} {Emulsions in porous media: From single
  droplet behavior to applications for oil recovery},}\ }\href
  {https://doi.org/https://doi.org/10.1016/j.cis.2018.03.002} {\bibfield
  {journal} {\bibinfo  {journal} {Advances in Colloid and Interface Science}\
  }\textbf {\bibinfo {volume} {256}},\ \bibinfo {pages} {305--325} (\bibinfo
  {year} {2018})}\BibitemShut {NoStop}%
\bibitem [{\citenamefont {Silva}\ \emph {et~al.}(2021)\citenamefont {Silva},
  \citenamefont {Coelho}, \citenamefont {da~Gama},\ and\ \citenamefont
  {Araújo}}]{arxiv.2101.06981}%
  \BibitemOpen
  \bibfield  {author} {\bibinfo {author} {\bibfnamefont {D.~P.~F.}\
  \bibnamefont {Silva}}, \bibinfo {author} {\bibfnamefont {R.~C.~V.}\
  \bibnamefont {Coelho}}, \bibinfo {author} {\bibfnamefont {M.~M.~T.}\
  \bibnamefont {da~Gama}},\ and\ \bibinfo {author} {\bibfnamefont {N.~A.~M.}\
  \bibnamefont {Araújo}},\ }\href {https://doi.org/10.48550/ARXIV.2101.06981}
  {\enquote {\bibinfo {title} {Effect of particle deformability on shear
  thinning in a 3d channel},}\ } (\bibinfo {year} {2021}),\ \bibinfo {note}
  {arxiv: 2101.06981}\BibitemShut {NoStop}%
\bibitem [{\citenamefont {Fei}\ \emph {et~al.}(2020)\citenamefont {Fei},
  \citenamefont {Scagliarini}, \citenamefont {Luo},\ and\ \citenamefont
  {Succi}}]{C9SM02331C}%
  \BibitemOpen
  \bibfield  {author} {\bibinfo {author} {\bibfnamefont {L.}~\bibnamefont
  {Fei}}, \bibinfo {author} {\bibfnamefont {A.}~\bibnamefont {Scagliarini}},
  \bibinfo {author} {\bibfnamefont {K.~H.}\ \bibnamefont {Luo}},\ and\ \bibinfo
  {author} {\bibfnamefont {S.}~\bibnamefont {Succi}},\ }\bibfield  {title}
  {\enquote {\bibinfo {title} {Discrete fluidization of dense monodisperse
  emulsions in neutral wetting microchannels},}\ }\href
  {https://doi.org/10.1039/C9SM02331C} {\bibfield  {journal} {\bibinfo
  {journal} {Soft Matter}\ }\textbf {\bibinfo {volume} {16}},\ \bibinfo {pages}
  {651--658} (\bibinfo {year} {2020})}\BibitemShut {NoStop}%
\bibitem [{\citenamefont {Foglino}\ \emph {et~al.}(2017)\citenamefont
  {Foglino}, \citenamefont {Morozov}, \citenamefont {Henrich},\ and\
  \citenamefont {Marenduzzo}}]{PhysRevLett.119.208002}%
  \BibitemOpen
  \bibfield  {author} {\bibinfo {author} {\bibfnamefont {M.}~\bibnamefont
  {Foglino}}, \bibinfo {author} {\bibfnamefont {A.~N.}\ \bibnamefont
  {Morozov}}, \bibinfo {author} {\bibfnamefont {O.}~\bibnamefont {Henrich}},\
  and\ \bibinfo {author} {\bibfnamefont {D.}~\bibnamefont {Marenduzzo}},\
  }\bibfield  {title} {\enquote {\bibinfo {title} {Flow of deformable droplets:
  Discontinuous shear thinning and velocity oscillations},}\ }\href
  {https://doi.org/10.1103/PhysRevLett.119.208002} {\bibfield  {journal}
  {\bibinfo  {journal} {Phys. Rev. Lett.}\ }\textbf {\bibinfo {volume} {119}},\
  \bibinfo {pages} {208002} (\bibinfo {year} {2017})}\BibitemShut {NoStop}%
\bibitem [{\citenamefont {Montessori}\ \emph
  {et~al.}(2021{\natexlab{a}})\citenamefont {Montessori}, \citenamefont
  {Tiribocchi}, \citenamefont {Lauricella}, \citenamefont {Bonaccorso},\ and\
  \citenamefont {Succi}}]{PhysRevFluids.6.023606}%
  \BibitemOpen
  \bibfield  {author} {\bibinfo {author} {\bibfnamefont {A.}~\bibnamefont
  {Montessori}}, \bibinfo {author} {\bibfnamefont {A.}~\bibnamefont
  {Tiribocchi}}, \bibinfo {author} {\bibfnamefont {M.}~\bibnamefont
  {Lauricella}}, \bibinfo {author} {\bibfnamefont {F.}~\bibnamefont
  {Bonaccorso}},\ and\ \bibinfo {author} {\bibfnamefont {S.}~\bibnamefont
  {Succi}},\ }\bibfield  {title} {\enquote {\bibinfo {title} {Wet to dry
  self-transitions in dense emulsions: From order to disorder and back},}\
  }\href {https://doi.org/10.1103/PhysRevFluids.6.023606} {\bibfield  {journal}
  {\bibinfo  {journal} {Phys. Rev. Fluids}\ }\textbf {\bibinfo {volume} {6}},\
  \bibinfo {pages} {023606} (\bibinfo {year} {2021}{\natexlab{a}})}\BibitemShut
  {NoStop}%
\bibitem [{\citenamefont {Montessori}\ \emph
  {et~al.}(2021{\natexlab{b}})\citenamefont {Montessori}, \citenamefont
  {Rocca}, \citenamefont {Prestininzi}, \citenamefont {Tiribocchi},\ and\
  \citenamefont {Succi}}]{doi:10.1063/5.0057501}%
  \BibitemOpen
  \bibfield  {author} {\bibinfo {author} {\bibfnamefont {A.}~\bibnamefont
  {Montessori}}, \bibinfo {author} {\bibfnamefont {M.~L.}\ \bibnamefont
  {Rocca}}, \bibinfo {author} {\bibfnamefont {P.}~\bibnamefont {Prestininzi}},
  \bibinfo {author} {\bibfnamefont {A.}~\bibnamefont {Tiribocchi}},\ and\
  \bibinfo {author} {\bibfnamefont {S.}~\bibnamefont {Succi}},\ }\bibfield
  {title} {\enquote {\bibinfo {title} {Deformation and breakup dynamics of
  droplets within a tapered channel},}\ }\href
  {https://doi.org/10.1063/5.0057501} {\bibfield  {journal} {\bibinfo
  {journal} {Physics of Fluids}\ }\textbf {\bibinfo {volume} {33}},\ \bibinfo
  {pages} {082008} (\bibinfo {year} {2021}{\natexlab{b}})}\BibitemShut
  {NoStop}%
\bibitem [{\citenamefont {Beatus}, \citenamefont {Tlusty},\ and\ \citenamefont
  {Bar-Ziv}(2006)}]{Beatus2006}%
  \BibitemOpen
  \bibfield  {author} {\bibinfo {author} {\bibfnamefont {T.}~\bibnamefont
  {Beatus}}, \bibinfo {author} {\bibfnamefont {T.}~\bibnamefont {Tlusty}},\
  and\ \bibinfo {author} {\bibfnamefont {R.}~\bibnamefont {Bar-Ziv}},\
  }\bibfield  {title} {\enquote {\bibinfo {title} {Phonons in a one-dimensional
  microfluidic crystal},}\ }\href {https://doi.org/10.1038/nphys432} {\bibfield
   {journal} {\bibinfo  {journal} {Nature Physics}\ }\textbf {\bibinfo {volume}
  {2}},\ \bibinfo {pages} {743--748} (\bibinfo {year} {2006})}\BibitemShut
  {NoStop}%
\bibitem [{\citenamefont {O’Connell}\ \emph {et~al.}(2019)\citenamefont
  {O’Connell}, \citenamefont {Lu}, \citenamefont {Browne},\ and\
  \citenamefont {Datta}}]{C9SM00300B}%
  \BibitemOpen
  \bibfield  {author} {\bibinfo {author} {\bibfnamefont {M.~G.}\ \bibnamefont
  {O’Connell}}, \bibinfo {author} {\bibfnamefont {N.~B.}\ \bibnamefont {Lu}},
  \bibinfo {author} {\bibfnamefont {C.~A.}\ \bibnamefont {Browne}},\ and\
  \bibinfo {author} {\bibfnamefont {S.~S.}\ \bibnamefont {Datta}},\ }\bibfield
  {title} {\enquote {\bibinfo {title} {Cooperative size sorting of deformable
  particles in porous media},}\ }\href {https://doi.org/10.1039/C9SM00300B}
  {\bibfield  {journal} {\bibinfo  {journal} {Soft Matter}\ }\textbf {\bibinfo
  {volume} {15}},\ \bibinfo {pages} {3620--3626} (\bibinfo {year}
  {2019})}\BibitemShut {NoStop}%
\bibitem [{\citenamefont {Zhao-Li}, \citenamefont {Chu-Guang},\ and\
  \citenamefont {Bao-Chang}(2002)}]{Zhao_Li_2002}%
  \BibitemOpen
  \bibfield  {author} {\bibinfo {author} {\bibfnamefont {G.}~\bibnamefont
  {Zhao-Li}}, \bibinfo {author} {\bibfnamefont {Z.}~\bibnamefont {Chu-Guang}},\
  and\ \bibinfo {author} {\bibfnamefont {S.}~\bibnamefont {Bao-Chang}},\
  }\bibfield  {title} {\enquote {\bibinfo {title} {Non-equilibrium
  extrapolation method for velocity and pressure boundary conditions in the
  lattice boltzmann method},}\ }\href
  {https://doi.org/10.1088/1009-1963/11/4/310} {\bibfield  {journal} {\bibinfo
  {journal} {Chinese Physics}\ }\textbf {\bibinfo {volume} {11}},\ \bibinfo
  {pages} {366--374} (\bibinfo {year} {2002})}\BibitemShut {NoStop}%
\bibitem [{\citenamefont {Kr\"{u}ger}\ \emph {et~al.}(2017)\citenamefont
  {Kr\"{u}ger}, \citenamefont {Kusumaatmaja}, \citenamefont {Kuzmin},
  \citenamefont {Shardt}, \citenamefont {Silva},\ and\ \citenamefont
  {Viggen}}]{kruger2016}%
  \BibitemOpen
  \bibfield  {author} {\bibinfo {author} {\bibfnamefont {T.}~\bibnamefont
  {Kr\"{u}ger}}, \bibinfo {author} {\bibfnamefont {H.}~\bibnamefont
  {Kusumaatmaja}}, \bibinfo {author} {\bibfnamefont {A.}~\bibnamefont
  {Kuzmin}}, \bibinfo {author} {\bibfnamefont {O.}~\bibnamefont {Shardt}},
  \bibinfo {author} {\bibfnamefont {G.}~\bibnamefont {Silva}},\ and\ \bibinfo
  {author} {\bibfnamefont {E.~M.}\ \bibnamefont {Viggen}},\ }\href
  {https://doi.org/10.1007/978-3-319-44649-3} {\emph {\bibinfo {title} {The
  Lattice Boltzmann Method}}}\ (\bibinfo  {publisher} {Springer International
  Publishing},\ \bibinfo {year} {2017})\BibitemShut {NoStop}%
\bibitem [{\citenamefont {Coelho}\ \emph {et~al.}(2021)\citenamefont {Coelho},
  \citenamefont {Moura}, \citenamefont {Telo~da Gama},\ and\ \citenamefont
  {Araújo}}]{https://doi.org/10.1002/fld.4988}%
  \BibitemOpen
  \bibfield  {author} {\bibinfo {author} {\bibfnamefont {R.~C.~V.}\
  \bibnamefont {Coelho}}, \bibinfo {author} {\bibfnamefont {C.~B.}\
  \bibnamefont {Moura}}, \bibinfo {author} {\bibfnamefont {M.~M.}\ \bibnamefont
  {Telo~da Gama}},\ and\ \bibinfo {author} {\bibfnamefont {N.~A.~M.}\
  \bibnamefont {Araújo}},\ }\bibfield  {title} {\enquote {\bibinfo {title}
  {Wetting boundary conditions for multicomponent pseudopotential lattice
  boltzmann},}\ }\href {https://doi.org/https://doi.org/10.1002/fld.4988}
  {\bibfield  {journal} {\bibinfo  {journal} {International Journal for
  Numerical Methods in Fluids}\ }\textbf {\bibinfo {volume} {93}},\ \bibinfo
  {pages} {2570--2580} (\bibinfo {year} {2021})}\BibitemShut {NoStop}%
\bibitem [{\citenamefont {Coelho}\ \emph {et~al.}(2022)\citenamefont {Coelho},
  \citenamefont {Cordeiro}, \citenamefont {Gazola},\ and\ \citenamefont
  {Teixeira}}]{Coelho_2022}%
  \BibitemOpen
  \bibfield  {author} {\bibinfo {author} {\bibfnamefont {R.~C.~V.}\
  \bibnamefont {Coelho}}, \bibinfo {author} {\bibfnamefont {L.~A. R.~G.}\
  \bibnamefont {Cordeiro}}, \bibinfo {author} {\bibfnamefont {R.~B.}\
  \bibnamefont {Gazola}},\ and\ \bibinfo {author} {\bibfnamefont {P.~I.~C.}\
  \bibnamefont {Teixeira}},\ }\bibfield  {title} {\enquote {\bibinfo {title}
  {Dynamics of two-dimensional liquid bridges},}\ }\href
  {https://doi.org/10.1088/1361-648x/ac594b} {\bibfield  {journal} {\bibinfo
  {journal} {Journal of Physics: Condensed Matter}\ }\textbf {\bibinfo {volume}
  {34}},\ \bibinfo {pages} {205001} (\bibinfo {year} {2022})}\BibitemShut
  {NoStop}%
\bibitem [{\citenamefont {Zuriguel}\ \emph {et~al.}(2014)\citenamefont
  {Zuriguel}, \citenamefont {Parisi}, \citenamefont {Hidalgo}, \citenamefont
  {Lozano}, \citenamefont {Janda}, \citenamefont {Gago}, \citenamefont
  {Peralta}, \citenamefont {Ferrer}, \citenamefont {Pugnaloni}, \citenamefont
  {Cl{\'{e}}ment}, \citenamefont {Maza}, \citenamefont {Pagonabarraga},\ and\
  \citenamefont {Garcimart{\'{\i}}n}}]{Zuriguel2014}%
  \BibitemOpen
  \bibfield  {author} {\bibinfo {author} {\bibfnamefont {I.}~\bibnamefont
  {Zuriguel}}, \bibinfo {author} {\bibfnamefont {D.~R.}\ \bibnamefont
  {Parisi}}, \bibinfo {author} {\bibfnamefont {R.~C.}\ \bibnamefont {Hidalgo}},
  \bibinfo {author} {\bibfnamefont {C.}~\bibnamefont {Lozano}}, \bibinfo
  {author} {\bibfnamefont {A.}~\bibnamefont {Janda}}, \bibinfo {author}
  {\bibfnamefont {P.~A.}\ \bibnamefont {Gago}}, \bibinfo {author}
  {\bibfnamefont {J.~P.}\ \bibnamefont {Peralta}}, \bibinfo {author}
  {\bibfnamefont {L.~M.}\ \bibnamefont {Ferrer}}, \bibinfo {author}
  {\bibfnamefont {L.~A.}\ \bibnamefont {Pugnaloni}}, \bibinfo {author}
  {\bibfnamefont {E.}~\bibnamefont {Cl{\'{e}}ment}}, \bibinfo {author}
  {\bibfnamefont {D.}~\bibnamefont {Maza}}, \bibinfo {author} {\bibfnamefont
  {I.}~\bibnamefont {Pagonabarraga}},\ and\ \bibinfo {author} {\bibfnamefont
  {A.}~\bibnamefont {Garcimart{\'{\i}}n}},\ }\bibfield  {title} {\enquote
  {\bibinfo {title} {Clogging transition of many-particle systems flowing
  through bottlenecks},}\ }\href {https://doi.org/10.1038/srep07324} {\bibfield
   {journal} {\bibinfo  {journal} {Scientific Reports}\ }\textbf {\bibinfo
  {volume} {4}} (\bibinfo {year} {2014}),\ 10.1038/srep07324}\BibitemShut
  {NoStop}%
\bibitem [{\citenamefont {Filho}\ \emph {et~al.}(2016)\citenamefont {Filho},
  \citenamefont {Moreira}, \citenamefont {Ara\'ujo}, \citenamefont {Andrade},\
  and\ \citenamefont {Herrmann}}]{PhysRevLett.117.275702}%
  \BibitemOpen
  \bibfield  {author} {\bibinfo {author} {\bibfnamefont {C.~I. N.~S.}\
  \bibnamefont {Filho}}, \bibinfo {author} {\bibfnamefont {A.~A.}\ \bibnamefont
  {Moreira}}, \bibinfo {author} {\bibfnamefont {N.~A.~M.}\ \bibnamefont
  {Ara\'ujo}}, \bibinfo {author} {\bibfnamefont {J.~S.}\ \bibnamefont
  {Andrade}},\ and\ \bibinfo {author} {\bibfnamefont {H.~J.}\ \bibnamefont
  {Herrmann}},\ }\bibfield  {title} {\enquote {\bibinfo {title} {Itinerant
  conductance in fuse-antifuse networks},}\ }\href
  {https://doi.org/10.1103/PhysRevLett.117.275702} {\bibfield  {journal}
  {\bibinfo  {journal} {Phys. Rev. Lett.}\ }\textbf {\bibinfo {volume} {117}},\
  \bibinfo {pages} {275702} (\bibinfo {year} {2016})}\BibitemShut {NoStop}%
\bibitem [{\citenamefont {Lam}, \citenamefont {Rosenbluth},\ and\ \citenamefont
  {Fletcher}(2008)}]{Lam2008}%
  \BibitemOpen
  \bibfield  {author} {\bibinfo {author} {\bibfnamefont {W.~A.}\ \bibnamefont
  {Lam}}, \bibinfo {author} {\bibfnamefont {M.~J.}\ \bibnamefont
  {Rosenbluth}},\ and\ \bibinfo {author} {\bibfnamefont {D.~A.}\ \bibnamefont
  {Fletcher}},\ }\bibfield  {title} {\enquote {\bibinfo {title} {Increased
  leukaemia cell stiffness is associated with symptoms of leucostasis in
  paediatric acute lymphoblastic leukaemia},}\ }\href
  {https://doi.org/10.1111/j.1365-2141.2008.07219.x} {\bibfield  {journal}
  {\bibinfo  {journal} {British Journal of Haematology}\ }\textbf {\bibinfo
  {volume} {142}},\ \bibinfo {pages} {497--501} (\bibinfo {year}
  {2008})}\BibitemShut {NoStop}%
\bibitem [{\citenamefont {Rosenbluth}, \citenamefont {Lam},\ and\ \citenamefont
  {Fletcher}(2008)}]{Rosenbluth2008}%
  \BibitemOpen
  \bibfield  {author} {\bibinfo {author} {\bibfnamefont {M.~J.}\ \bibnamefont
  {Rosenbluth}}, \bibinfo {author} {\bibfnamefont {W.~A.}\ \bibnamefont
  {Lam}},\ and\ \bibinfo {author} {\bibfnamefont {D.~A.}\ \bibnamefont
  {Fletcher}},\ }\bibfield  {title} {\enquote {\bibinfo {title} {Analyzing cell
  mechanics in hematologic diseases with microfluidic biophysical flow
  cytometry},}\ }\href {https://doi.org/10.1039/b802931h} {\bibfield  {journal}
  {\bibinfo  {journal} {Lab on a Chip}\ }\textbf {\bibinfo {volume} {8}},\
  \bibinfo {pages} {1062} (\bibinfo {year} {2008})}\BibitemShut {NoStop}%
\bibitem [{\citenamefont {Wong}\ \emph {et~al.}(2002)\citenamefont {Wong},
  \citenamefont {Song}, \citenamefont {Grimes}, \citenamefont {Fu},
  \citenamefont {Dewhirst}, \citenamefont {Muschel},\ and\ \citenamefont
  {Al-Mehdi}}]{Wong2002}%
  \BibitemOpen
  \bibfield  {author} {\bibinfo {author} {\bibfnamefont {C.~W.}\ \bibnamefont
  {Wong}}, \bibinfo {author} {\bibfnamefont {C.}~\bibnamefont {Song}}, \bibinfo
  {author} {\bibfnamefont {M.~M.}\ \bibnamefont {Grimes}}, \bibinfo {author}
  {\bibfnamefont {W.}~\bibnamefont {Fu}}, \bibinfo {author} {\bibfnamefont
  {M.~W.}\ \bibnamefont {Dewhirst}}, \bibinfo {author} {\bibfnamefont {R.~J.}\
  \bibnamefont {Muschel}},\ and\ \bibinfo {author} {\bibfnamefont {A.-B.}\
  \bibnamefont {Al-Mehdi}},\ }\bibfield  {title} {\enquote {\bibinfo {title}
  {Intravascular location of breast cancer cells after spontaneous metastasis
  to the lung},}\ }\href {https://doi.org/10.1016/s0002-9440(10)64233-2}
  {\bibfield  {journal} {\bibinfo  {journal} {The American Journal of
  Pathology}\ }\textbf {\bibinfo {volume} {161}},\ \bibinfo {pages} {749--753}
  (\bibinfo {year} {2002})}\BibitemShut {NoStop}%
\bibitem [{\citenamefont {Preira}\ \emph {et~al.}(2012)\citenamefont {Preira},
  \citenamefont {Leoni}, \citenamefont {Valignat}, \citenamefont {Lellouch},
  \citenamefont {Robert}, \citenamefont {Forel}, \citenamefont {Papazian},
  \citenamefont {Dumenil}, \citenamefont {Bongrand},\ and\ \citenamefont
  {Th\'{e}odoly}}]{doi:10.1504/IJNT.2012.045340}%
  \BibitemOpen
  \bibfield  {author} {\bibinfo {author} {\bibfnamefont {P.}~\bibnamefont
  {Preira}}, \bibinfo {author} {\bibfnamefont {T.}~\bibnamefont {Leoni}},
  \bibinfo {author} {\bibfnamefont {M.}~\bibnamefont {Valignat}}, \bibinfo
  {author} {\bibfnamefont {A.}~\bibnamefont {Lellouch}}, \bibinfo {author}
  {\bibfnamefont {P.}~\bibnamefont {Robert}}, \bibinfo {author} {\bibfnamefont
  {J.}~\bibnamefont {Forel}}, \bibinfo {author} {\bibfnamefont
  {L.}~\bibnamefont {Papazian}}, \bibinfo {author} {\bibfnamefont
  {G.}~\bibnamefont {Dumenil}}, \bibinfo {author} {\bibfnamefont
  {P.}~\bibnamefont {Bongrand}},\ and\ \bibinfo {author} {\bibfnamefont
  {O.}~\bibnamefont {Th\'{e}odoly}},\ }\bibfield  {title} {\enquote {\bibinfo
  {title} {Microfluidic tools to investigate pathologies in the blood
  microcirculation},}\ }\href {https://doi.org/10.1504/IJNT.2012.045340}
  {\bibfield  {journal} {\bibinfo  {journal} {International Journal of
  Nanotechnology}\ }\textbf {\bibinfo {volume} {9}},\ \bibinfo {pages}
  {529--547} (\bibinfo {year} {2012})}\BibitemShut {NoStop}%
\bibitem [{\citenamefont {Hogg}(1987)}]{Hogg1987}%
  \BibitemOpen
  \bibfield  {author} {\bibinfo {author} {\bibfnamefont {J.~C.}\ \bibnamefont
  {Hogg}},\ }\bibfield  {title} {\enquote {\bibinfo {title} {Neutrophil
  kinetics and lung injury},}\ }\href
  {https://doi.org/10.1152/physrev.1987.67.4.1249} {\bibfield  {journal}
  {\bibinfo  {journal} {Physiological Reviews}\ }\textbf {\bibinfo {volume}
  {67}},\ \bibinfo {pages} {1249--1295} (\bibinfo {year} {1987})}\BibitemShut
  {NoStop}%
\bibitem [{\citenamefont {Hotchkiss}\ and\ \citenamefont
  {Karl}(2003)}]{Hotchkiss2003}%
  \BibitemOpen
  \bibfield  {author} {\bibinfo {author} {\bibfnamefont {R.~S.}\ \bibnamefont
  {Hotchkiss}}\ and\ \bibinfo {author} {\bibfnamefont {I.~E.}\ \bibnamefont
  {Karl}},\ }\bibfield  {title} {\enquote {\bibinfo {title} {The
  pathophysiology and treatment of sepsis},}\ }\href
  {https://doi.org/10.1056/nejmra021333} {\bibfield  {journal} {\bibinfo
  {journal} {New England Journal of Medicine}\ }\textbf {\bibinfo {volume}
  {348}},\ \bibinfo {pages} {138--150} (\bibinfo {year} {2003})}\BibitemShut
  {NoStop}%
\bibitem [{\citenamefont {Yoshida}\ \emph {et~al.}(2006)\citenamefont
  {Yoshida}, \citenamefont {Kondo}, \citenamefont {Wang},\ and\ \citenamefont
  {Doerschuk}}]{Yoshida2006}%
  \BibitemOpen
  \bibfield  {author} {\bibinfo {author} {\bibfnamefont {K.}~\bibnamefont
  {Yoshida}}, \bibinfo {author} {\bibfnamefont {R.}~\bibnamefont {Kondo}},
  \bibinfo {author} {\bibfnamefont {Q.}~\bibnamefont {Wang}},\ and\ \bibinfo
  {author} {\bibfnamefont {C.~M.}\ \bibnamefont {Doerschuk}},\ }\bibfield
  {title} {\enquote {\bibinfo {title} {Neutrophil cytoskeletal rearrangements
  during capillary sequestration in bacterial pneumonia in rats},}\ }\href
  {https://doi.org/10.1164/rccm.200502-276oc} {\bibfield  {journal} {\bibinfo
  {journal} {American Journal of Respiratory and Critical Care Medicine}\
  }\textbf {\bibinfo {volume} {174}},\ \bibinfo {pages} {689--698} (\bibinfo
  {year} {2006})}\BibitemShut {NoStop}%
\bibitem [{\citenamefont {Preira}\ \emph {et~al.}(2013)\citenamefont {Preira},
  \citenamefont {Grandné}, \citenamefont {Forel}, \citenamefont {Gabriele},
  \citenamefont {Camara},\ and\ \citenamefont {Theodoly}}]{C2LC40847C}%
  \BibitemOpen
  \bibfield  {author} {\bibinfo {author} {\bibfnamefont {P.}~\bibnamefont
  {Preira}}, \bibinfo {author} {\bibfnamefont {V.}~\bibnamefont {Grandné}},
  \bibinfo {author} {\bibfnamefont {J.-M.}\ \bibnamefont {Forel}}, \bibinfo
  {author} {\bibfnamefont {S.}~\bibnamefont {Gabriele}}, \bibinfo {author}
  {\bibfnamefont {M.}~\bibnamefont {Camara}},\ and\ \bibinfo {author}
  {\bibfnamefont {O.}~\bibnamefont {Theodoly}},\ }\bibfield  {title} {\enquote
  {\bibinfo {title} {Passive circulating cell sorting by deformability using a
  microfluidic gradual filter},}\ }\href {https://doi.org/10.1039/C2LC40847C}
  {\bibfield  {journal} {\bibinfo  {journal} {Lab Chip}\ }\textbf {\bibinfo
  {volume} {13}},\ \bibinfo {pages} {161--170} (\bibinfo {year}
  {2013})}\BibitemShut {NoStop}%
\bibitem [{\citenamefont {Guo}, \citenamefont {Zheng},\ and\ \citenamefont
  {Shi}(2002)}]{guo_discrite_2002}%
  \BibitemOpen
  \bibfield  {author} {\bibinfo {author} {\bibfnamefont {Z.}~\bibnamefont
  {Guo}}, \bibinfo {author} {\bibfnamefont {C.}~\bibnamefont {Zheng}},\ and\
  \bibinfo {author} {\bibfnamefont {B.}~\bibnamefont {Shi}},\ }\bibfield
  {title} {\enquote {\bibinfo {title} {Discrete lattice effects on the forcing
  term in the lattice boltzmann method},}\ }\href
  {https://doi.org/10.1103/PhysRevE.65.046308} {\bibfield  {journal} {\bibinfo
  {journal} {Phys. Rev. E}\ }\textbf {\bibinfo {volume} {65}},\ \bibinfo
  {pages} {046308} (\bibinfo {year} {2002})}\BibitemShut {NoStop}%
\bibitem [{\citenamefont {Chikatamarla}\ and\ \citenamefont
  {Karlin}(2009)}]{Chikatamarla_2009}%
  \BibitemOpen
  \bibfield  {author} {\bibinfo {author} {\bibfnamefont {S.~S.}\ \bibnamefont
  {Chikatamarla}}\ and\ \bibinfo {author} {\bibfnamefont {I.~V.}\ \bibnamefont
  {Karlin}},\ }\bibfield  {title} {\enquote {\bibinfo {title} {Lattices for the
  lattice boltzmann method},}\ }\href
  {https://doi.org/10.1103/PhysRevE.79.046701} {\bibfield  {journal} {\bibinfo
  {journal} {Phys. Rev. E}\ }\textbf {\bibinfo {volume} {79}},\ \bibinfo
  {pages} {046701} (\bibinfo {year} {2009})}\BibitemShut {NoStop}%
\bibitem [{\citenamefont {Benzi}, \citenamefont {Chibbaro},\ and\ \citenamefont
  {Succi}(2009)}]{PhysRevLett.102.026002}%
  \BibitemOpen
  \bibfield  {author} {\bibinfo {author} {\bibfnamefont {R.}~\bibnamefont
  {Benzi}}, \bibinfo {author} {\bibfnamefont {S.}~\bibnamefont {Chibbaro}},\
  and\ \bibinfo {author} {\bibfnamefont {S.}~\bibnamefont {Succi}},\ }\bibfield
   {title} {\enquote {\bibinfo {title} {Mesoscopic lattice boltzmann modeling
  of flowing soft systems},}\ }\href
  {https://doi.org/10.1103/PhysRevLett.102.026002} {\bibfield  {journal}
  {\bibinfo  {journal} {Phys. Rev. Lett.}\ }\textbf {\bibinfo {volume} {102}},\
  \bibinfo {pages} {026002} (\bibinfo {year} {2009})}\BibitemShut {NoStop}%
\bibitem [{\citenamefont {Tiribocchi}\ \emph {et~al.}(2021)\citenamefont
  {Tiribocchi}, \citenamefont {Montessori}, \citenamefont {Lauricella},
  \citenamefont {Bonaccorso}, \citenamefont {Succi}, \citenamefont {Aime},
  \citenamefont {Milani},\ and\ \citenamefont {Weitz}}]{Tiribocchi2021}%
  \BibitemOpen
  \bibfield  {author} {\bibinfo {author} {\bibfnamefont {A.}~\bibnamefont
  {Tiribocchi}}, \bibinfo {author} {\bibfnamefont {A.}~\bibnamefont
  {Montessori}}, \bibinfo {author} {\bibfnamefont {M.}~\bibnamefont
  {Lauricella}}, \bibinfo {author} {\bibfnamefont {F.}~\bibnamefont
  {Bonaccorso}}, \bibinfo {author} {\bibfnamefont {S.}~\bibnamefont {Succi}},
  \bibinfo {author} {\bibfnamefont {S.}~\bibnamefont {Aime}}, \bibinfo {author}
  {\bibfnamefont {M.}~\bibnamefont {Milani}},\ and\ \bibinfo {author}
  {\bibfnamefont {D.~A.}\ \bibnamefont {Weitz}},\ }\bibfield  {title} {\enquote
  {\bibinfo {title} {The vortex-driven dynamics of droplets within droplets},}\
  }\href {https://doi.org/10.1038/s41467-020-20364-0} {\bibfield  {journal}
  {\bibinfo  {journal} {Nature Communications}\ }\textbf {\bibinfo {volume}
  {12}} (\bibinfo {year} {2021}),\ 10.1038/s41467-020-20364-0}\BibitemShut
  {NoStop}%
\end{thebibliography}%

\end{document}